\documentclass[prb,twocolumn,amsmath,amsfonts,superscriptaddress,floatfix]{revtex4-1}

\usepackage{amssymb}
\usepackage[mathscr]{eucal}
\usepackage[dvips]{graphicx}
\usepackage{color}
\usepackage{bbold}

\newcommand{\tr}{\textrm{tr}}

\newcommand{\ket}[1]{| \, #1 \, \rangle}
\newcommand{\bra}[1]{\langle \, #1 \, |}

\begin{document}

\title{\fontsize{12pt}{12pt} \selectfont The Algebraic Bethe Ansatz and Tensor Networks}

\author{V. Murg}
\affiliation{\mbox{Vienna Center for Quantum Science and Technology, Faculty of Physics, University of Vienna, Vienna, Austria}}

\author{V. E. Korepin}
\affiliation{C. N. Yang Institute for Theoretical Physics, State University of New York at Stony Brook, NY 11794-3840, USA}

\author{F. Verstraete}
\affiliation{\mbox{Vienna Center for Quantum Science and Technology, Faculty of Physics, University of Vienna, Vienna, Austria}}

\date{\today}

\begin{abstract}
We describe the Algebraic Bethe Ansatz 
for the spin~$1/2$ XXX and XXZ Heisenberg chains with open and periodic boundary conditions
in terms of tensor networks.
These Bethe eigenstates
have the structure of Matrix Product States with a conserved number of down-spins.
The tensor network formulation suggestes
possible extensions of the Algebraic Bethe Ansatz to two dimensions.
\end{abstract}


\maketitle


\section{Introduction}

The coordinate Bethe ansatz~\cite{bethe31} is an extremely successful method for solving
one-dimensional problems exactly. 
It reduces the complex problem of diagonalizing the Hamiltonian to finding the solutions of a set of algebraic equations.
Once solutions to these algebraic equations are found -- numerical approaches to find them efficiently exist in many cases -- 
the eigenvalues are known exactly. However, the eigenstates are available only as a complex mathematical expressions
the structure of which is not evident.
This makes it insuperable, in general, to get interesting properties out of the states -- like their entanglement characteristics or
their correlations.
The algebraic Bethe ansatz~\cite{korepin93} reveals more about the structure of the eigenstate
and offers new perspectives to obtain scalar products~\cite{slavnov07}, norms~\cite{korepin93} and correlations~\cite{korepin93}.

In this paper, we point out this structure by formulating the algebraic Bethe ansatz in the pictoresque tensor network language.
In addition to making the ansatz more vivid, the tensor network formulation might
bear the potential of extending the ansatz to higher dimensions.

The description of states in terms of tensor networks has been very successful in the recent past.
The one-dimensional matrix product states (MPS)~\cite{affleck87,affleck88} form the basis for the extremely successful
density matrix normalization group (DMRG)~\cite{white92,white92b}.
Also, they have attracted considerable interest in the interdisciplinary field of quantum information
and condensed matter physics~\cite{verstraetecirac05,perezgarcia06,verstraeteciracmurg08,singh10}.
For describing the ground state of systems on higher-dimensional lattices,
the projected entangled pair states (PEPS)~\cite{verstraetecirac04} were introduced and
proved to be useful for the numerical study of ground states of two-dimensional systems~\cite{murgverstraete07,murgverstraete08}.
The Multiscale Entanglement Renormalization Ansatz (MERA)~\cite{vidal05,vidal06}
allows the description and numerical study of critical systems.

From the tensor network desription of the Bethe eigenstates
it is immediately obvious that eigenstates can be described as MPS:
see also Katsura and Maruyama~[\onlinecite{katsura10}].
Katsura and Maruyama also show that the alternative formulation of the Bethe Ansatz 
by Alcaraz and Lazo~[\onlinecite{alcaraz04,alcaraz04b,alcaraz06}]
is equivalent to the algebraic Bethe ansatz.

In Sec.~\ref{sec:mpsandbethe}, we describe the tensor network form
of the Bethe eigenstates and the structure of the obtained MPS.
In Sec.~\ref{sec:bethe}, we formulate the algebraic Bethe ansatz
in the tensor network language.
In Sec.\ref{sec:openbc}, we give a pictoresque description of the algebraic
Bethe ansatz with open boundary conditions in terms of tensor networks.


\section{Matrix Product State form of Bethe Solutions} \label{sec:mpsandbethe}

\begin{figure}[t]
    \begin{center}
        \includegraphics[width=0.48\textwidth]{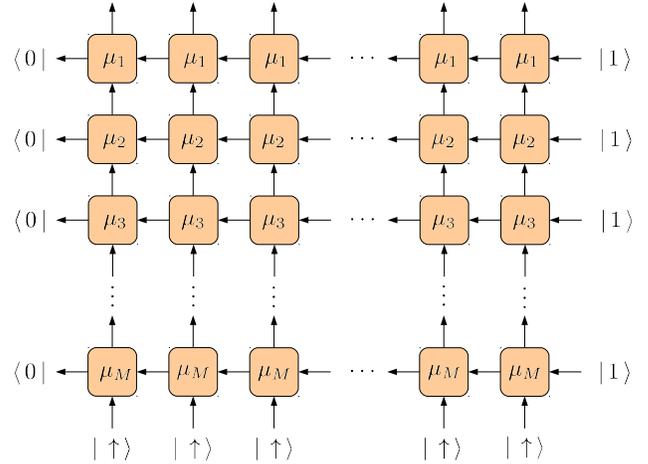}
     \end{center}
    \caption{
	Tensor network constituting the Bethe eigenstate of the
	Heisenberg model or XXZ model with periodic boundary conditions.
       }
    \label{fig:betheansatz}
\end{figure}

Typically, Bethe-eigenstates are obtained as
products of operators $B(\mu_j)$ applied on a certain vacuum state $\ket{vac}$, i.e.
\begin{equation} \label{eqn:betheansatz}
\ket{\Psi(\mu_1,\ldots,\mu_M)} = B(\mu_1) \cdots B(\mu_M) \ket{vac}.
\end{equation}
The parameters $\{\mu_j\}$ are thereby solutions of Bethe equations
and the $B(\mu_j)$'s play the role of creation operators.
In case of the antiferromagnetic Heisenberg model and the XXZ model with periodic boundary conditions,
the vacuum corresponds to the state with all spins up
and each operator $B(\mu_j)$ creates one down-spin.
Thus, the product of $M$ such operators applied to the vacuum creates a state with $M$ down-spins, i.e. magnetization $S_z = N/2 - M$
(with $N$ being the number of spins).
$B(\lambda)$ is an operator acting on the whole Hilbert-space of dimension $2^N$, but it has the
well-structured form of a Matrix Product Operator (MPO)~\cite{verstraeteripoll04} with virtual dimension~$2$.
Indeed, as will be shown in Sec.~\ref{sec:bethe},
\begin{displaymath}
B(\lambda)
=
\sum_{\begin{smallmatrix}k_1 \cdots k_N\\l_1 \cdots l_N\end{smallmatrix} }
\bra{0}
\mathcal{L}^{k_1}_{l_1}(\lambda)
\cdots
\mathcal{L}^{k_N}_{l_N}(\lambda)
\ket{1}
o^{k_1}_{l_1} \otimes \cdots \otimes o^{k_N}_{l_N}
\end{displaymath}
with $k,l \in \{0,1\}$, $o^k_l = \ket{k}\bra{l}$ ($0 \equiv \uparrow$, $1 \equiv \downarrow$) 
and $\mathcal{L}^{k}_{l}(\lambda)$ being $2 \times 2$
matrices dependent on the parameter~$\lambda$.
The product of operators $B(\mu_1) \cdots B(\mu_M)$ can be read as
the contraction of the set of $4$-index tensors $[\mathcal{L}^{k}_{l}(\mu_j)]^{r}_{r'}$
with respect to a rectangular grid, as shown in Fig.~\ref{fig:betheansatz}.
Thereby, $r$, $r'$, $k$ and $l$ label the left, right, up and down-indices, respectively.
Explicitely, the matrices $\mathcal{L}^{k}_{l}(\lambda)$ read
\begin{displaymath}
\begin{array}{cc}
\mathcal{L}^{0}_{0}(\lambda) = \left( \begin{array}{cc} 1 & 0 \\ 0 & c(\lambda) \end{array} \right), &
\mathcal{L}^{0}_{1}(\lambda) = \left( \begin{array}{cc} 0 & 0 \\ b(\lambda) & 0 \end{array} \right) \\ \\
\mathcal{L}^{1}_{0}(\lambda) = \left( \begin{array}{cc} 0 & b(\lambda) \\ 0 & 0 \end{array} \right), &
\mathcal{L}^{1}_{1}(\lambda) = \left( \begin{array}{cc} c(\lambda) & 0 \\ 0 & 1 \end{array} \right).
\end{array}
\end{displaymath}
In case of the Heisenberg model $H_{XXX} = \sum_{j=1}^N h_{XXX}^{(j,j+1)}$ with
\begin{displaymath}
h_{XXX} = \frac{1}{2} \left[ \sigma_x \otimes \sigma_x + \sigma_y \otimes \sigma_y + 
\sigma_z \otimes \sigma_z - \mathbb{1} \right],
\end{displaymath}
the functions $b(\lambda)$ and $c(\lambda)$ are
\begin{displaymath}
b(\lambda) = \frac{1}{1+\lambda}, \hspace{0.5cm}
c(\lambda) = \frac{\lambda}{1+\lambda}.
\end{displaymath}
For the XXZ model $H_{XXZ}(\Delta) = \sum_{j=1}^N h_{XXZ}^{(j,j+1)}(\Delta)$ with
\begin{displaymath}
h_{XXZ}(\Delta) = \frac{1}{2} \left[ \sigma_x \otimes \sigma_x + \sigma_y \otimes \sigma_y + 
\Delta \left( \sigma_z \otimes \sigma_z - \mathbb{1} \right) \right],
\end{displaymath}
the functions read
\begin{displaymath}
b(\lambda) = \frac{\sinh(2 i \eta)}{\sinh(\lambda+2 i \eta)}, \hspace{0.5cm}
c(\lambda) = \frac{\sinh(\lambda)}{\sinh(\lambda+2 i \eta)}.
\end{displaymath}
The parameter $\eta$ is related to the inhomogenity $\Delta$ in the XXZ model via $\Delta = \cos(2\eta)$.

Because of its ``creation operator''-property, there is an inherent structure in the MPO~$B(\mu)$:
each summand in the MPO $B(\mu)$ must be non-zero only if $k_1+\cdots+k_N = l_1+\cdots+l_N + 1$.
This global constraint can be reduced to the local constraint
that the tensors $[\mathcal{L}^{k}_{l}(\mu)]^{r}_{r'}$ must be non-zero only if
$r' = r + (k-l)$. This allows to interprete the virtual indices as ``creation-annihilation'' counters:
the right index $r'$ is equal to the left index $r$ if the physical state is unchanged, it is
increased if a down-spin is created and is decreased if a down-spin is annihilated.
Thus, the virtual indices transfer the information on how many down-spins are created and anniliated from
left to right. Since the left boundary-state is $\bra{0}$ and the right boundary-state is $\ket{1}$,
it is guaranteed that the whole MPO creates exactly one down-spin.
With the restriction $r,r' \in \{0,1\}$ there are $6$ possible configurations that fulfill the local
constraint. In other words, only $6$ entries of the tensor $[\mathcal{L}^{k}_{l}(\mu)]^{r}_{r'}$
are non-zero. These $6$ non-zero entries are
\begin{displaymath}
\begin{array}{ccccc}
\small[\mathcal{L}^{0}_{0}(\mu)]^{0}_{0}, & & [\mathcal{L}^{1}_{1}(\mu)]^{1}_{1} \\
\small[\mathcal{L}^{0}_{1}(\mu)]^{1}_{0}, & & [\mathcal{L}^{1}_{0}(\mu)]^{0}_{1} \\
\small[\mathcal{L}^{0}_{0}(\mu)]^{1}_{1}, & & [\mathcal{L}^{1}_{1}(\mu)]^{0}_{0}, \\
\end{array}
\end{displaymath}
which is consistent with the matrices written above.

The multiplication of all MPOs with the product state~$\ket{vac}$ evidently 
yields a Matrix Product State (MPS)~\cite{verstraeteciracmurg08,verstraetecirac05}
with bond-dimension~$2^M$.
Since each MPO $B(\lambda)$ has the ``creation operator''-property to
create one down-spin, the MPS contains exactly~$M$ down-spins.
Explicitly, the MPS reads
\begin{displaymath}
\ket{\Psi} = \sum_{k_1 \cdots k_N} \bra{0} \bra{0} \mathcal{A}^{k_1} \cdots \mathcal{A}^{k_N} \ket{0} \ket{M} \ket{k_1,\ldots,k_N}
\end{displaymath}
with matrices $\mathcal{A}^k$ being block-diagonal in the sense that
$\bra{\alpha} \bra{s} \mathcal{A}^k \ket{\beta} \ket{s'} \equiv [\mathcal{A}^k]^{\alpha s}_{\beta s'}$.
$\alpha$ and $\beta$ are the virtual indices that range from $0$ to $D-1$ (with $D$ being the virtual dimension of the state).
One the other hand, $s$ and $s'$ are the symmetry indices that transfer the information about the number
of down-spins from left to right. The local constraint that guarantees this information transfer is
$s' = s + k$. This constraint determines the blocks $[\mathcal{A}^k]^{- s}_{- s'}$ that are non-zero
and allows a sparse storage of the state.
The left boundary-state $\bra{0}$ and the right boundary-state $\ket{M}$ fix the
total number of down-spins of the MPS to~$M$.

\begin{table}
\begin{tabular}{l|l|rcl}
m & D &   & &    \\
\hline
$1$ & $2$ & $2$ & $=$ & $1 \oplus 1$ \\
$2$ & $4$ & $2 \otimes 2$ & $=$ & $1 \oplus 2 \oplus 1$ \\
$3$ & $8$ & $2 \otimes 2 \otimes 2$ & $=$ & $1 \oplus 4 \oplus 2 \oplus 1$ \\
$4$ & $16$ & $2 \otimes 2 \otimes 2 \otimes 2$ & $=$ & $1 \oplus 8 \oplus 4 \oplus 2 \oplus 1$ \\
$5$ & $32$ & $2 \otimes 2 \otimes 2 \otimes 2 \otimes 2$ & $=$ & $1 \oplus 16 \oplus 8 \oplus 4 \oplus 2 \oplus 1$ \\
$6$ & $64$ & $2 \otimes 2 \otimes 2 \otimes 2 \otimes 2 \otimes 2$ & $=$ & $1 \oplus 32 \oplus 16 \oplus 8 \oplus 4 \oplus 2 \oplus 1$ \\
\end{tabular}
\caption{
	Disintegration of the matrices forming the MPS at step~$m$, $\ket{\Psi_m}$, into blocks.
	The full size of the matrices is $D \times D$.
	The MPS has a conserved number of $m$ down-spins.
}
\label{tab:blocks}
\end{table}

The MPS is constructed iteratively by applying the MPOs $B(\mu_1),\ldots,B(\mu_M)$ successively to the vacuum state~$\ket{vac}$.
The state after~$m$ multiplications is evidently a MPS with $m$ down-spins
which shall be denoted as
\begin{displaymath}
\ket{\Psi_m} = \sum_{k_1 \cdots k_N} \bra{0} \bra{0} \mathcal{A}^{k_1}_m \cdots \mathcal{A}^{k_N}_m \ket{0} \ket{m} \ket{k_1,\ldots,k_N}
\end{displaymath}
with
$\mathcal{A}^k_m$ being block-diagonal in the sense that
$\bra{\alpha} \bra{s} \mathcal{A}^k_m \ket{\beta} \ket{s'} \equiv [\mathcal{A}^k_m]^{\alpha s}_{\beta s'}$
and fulfilling the constraint $s'=s+k$, as before.
The application of the operator $B(\mu)$ to $\ket{\Psi_m}$ yields a state 
with $m+1$ down-spins
\begin{displaymath}
\ket{\Psi_{m+1}} = \sum_{k_1 \cdots k_N} \bra{0} \bra{0} \mathcal{A}^{k_1}_{m+1} \cdot \cdot \mathcal{A}^{k_N}_{m+1} \ket{0} \ket{m+1} \ket{k_1,..,k_N}.
\end{displaymath}
The matrices $\mathcal{A}^{k}_{m+1}$ emerge from tensor-products between $\mathcal{L}^{k}_{l}$ and $\mathcal{A}^l_m$,
i.e. $\mathcal{A}^{k}_{m+1} = \sum_l \mathcal{L}^{k}_{l} \otimes \mathcal{A}^l_m$.
In index notation,
\begin{displaymath}
\bra{\alpha} \bra{r} \bra{s} \mathcal{A}^{k}_{m+1} \ket{\beta} \ket{r'} \ket{s'} =
\sum_l \bra{r} \mathcal{L}^{k}_{l} \ket{r'}
\bra{\alpha} \bra{s} \mathcal{A}^l \ket{\beta} \ket{s'}.
\end{displaymath}
Because of the constraints $s'=s+l$ and $r'=r+(k-l)$,
$S=s+r$ and $S'=s'+r'$ suggest themselves as new symmetry indices.
With this definition, $S'=S+k$, as desired.
$S$ and $S'$ range from $0$ to $m+1$, since $s \in \{0,\ldots,m\}$ and $r \in \{0,1\}$.
For $S=0$ and $S=m+1$, there is the unique choice for $s=r=0$ and $s=m, r=1$, respectively.
For $0<S<m+1$, either $s=S,r=0$, or $s=S-1,r=1$.
In this case, the index $r$ must be kept to resolve this ambiguity.
The index $r$ can be incorporated into a new virtual index $\tilde{\alpha}$ as $\tilde{\alpha} = (\alpha, r)$.
Thus, the dimension of the blocks doubles for $0<S<m+1$.
The column indices $S'$, $s'$, $r'$ and $\beta$ can be treated in the same way:
for $S'=0$ and $S'=m+1$, $s'$ and $r'$ are unambiguously defined;
for $0<S'<m+1$ there is an ambiguity that can to be resolved by
incorporating index $r'$ into a new virtual index $\tilde{\beta} = (\beta,r')$.
The matrices $\mathcal{A}^{k}_{m+1}$ in terms of the virtual indices $\tilde{\alpha}$ and $\tilde{\beta}$ and
the symmetry indices $S$ and $S'$, i.e.
\begin{displaymath}
\bra{\tilde{\alpha}} \bra{S} \mathcal{A}^{k}_{m+1} \ket{\tilde{\beta}} \ket{S'}
:=
\bra{\alpha} \bra{r} \bra{s} \mathcal{A}^{k}_{m+1} \ket{\beta} \ket{r'} \ket{s'},
\end{displaymath}
have the desired block-form that fulfills the constraint $S'=S+k$.
Please refer to Table~\ref{tab:blocks} to see the dimensions of the block-representations
that arise for different $m$'s.

\begin{figure}[t]
    \begin{center}
        \includegraphics[width=0.48\textwidth]{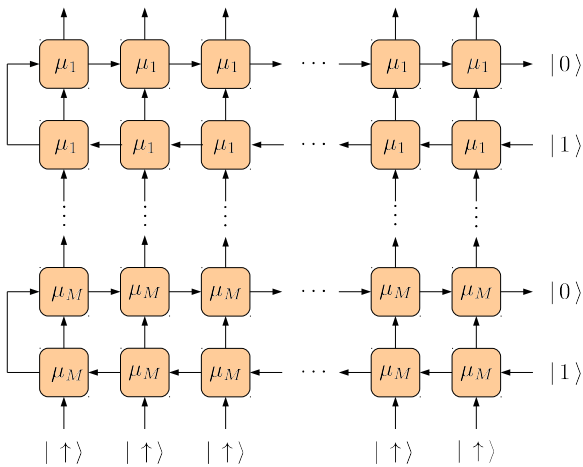}
     \end{center}
    \caption{
	Tensor network constituting the Bethe eigenstate of the
	Heisenberg model or XXZ model with open boundary conditions.
       }
    \label{fig:betheansatzopenbc}
\end{figure}

In the case of open boundary conditions, the Bethe Ansatz has the same form as in~$(\ref{eqn:betheansatz})$,
merely the creation Operators are not single MPOs, but products of two MPOs~\cite{cherednik84,sklyanin88}:
\begin{displaymath}
\mathcal{B}(\mu) = \sum_{s=0}^1 \bar{B}_s(\mu) B_{1-s}(\mu)
\end{displaymath}
$B_{1-s}(\mu)$ has the property to create $1-s$ down-spins, whereas $\bar{B}_{s}(\mu)$ creates $s$ down-spins ($s \in \{0,1\}$),
such that $\mathcal{B}(\mu)$ is a creation operator for exactly one down-spin, as before.
In terms of the previously defined $2 \times 2$ matrices $\mathcal{L}^{k}_{l}(\mu)$, the MPOs read
(see Sec.~\ref{sec:openbc})
\begin{displaymath}
B_s(\mu)
=
\sum_{\begin{smallmatrix}k_1 \cdots k_N\\l_1 \cdots l_N\end{smallmatrix} }
\bra{s}
\mathcal{L}^{k_1}_{l_1}(\mu)
\cdots
\mathcal{L}^{k_N}_{l_N}(\mu)
\ket{1}
o^{k_1}_{l_1} \otimes \cdots \otimes o^{k_N}_{l_N}
\end{displaymath}
and
\begin{displaymath}
\bar{B}_{1-s}(\mu)
=
\sum_{\begin{smallmatrix}k_1 \cdots k_N\\l_1 \cdots l_N\end{smallmatrix} }
\bra{s}
\mathcal{L}^{k_1}_{l_1}(\mu)^T
\cdots
\mathcal{L}^{k_N}_{l_N}(\mu)^T
\ket{0}
o^{k_1}_{l_1} \otimes \cdots \otimes o^{k_N}_{l_N}.
\end{displaymath}
The virtual indices of $B_s(\mu)$ indicate the balance of created versus annihilated down-spins
from left to right. 
This is due to the local constraint on $[\mathcal{L}^{k}_{l}(\mu)]^{r}_{r'}$ that $r' = r + (k-l)$,
as mentioned before.
Since the left boundary-vector is~$\bra{0}$ and the right boundary-vector is~$\ket{s}$,
the creation of $s$ down-spins is guaranteed.
In case of $\bar{B}_{1-s}(\mu)$, the MPO is built from the transposed matrices $\mathcal{L}^{k}_{l}(\mu)^T$,
such that the local constraint on $[\mathcal{L}^{k}_{l}(\mu)^T]^{r}_{r'}$ is $r = r' + (k-1)$ and
the virtual indices count the creation-annihilation balance from right to left.
With the right boundary vector~$\ket{s}$ and the left boundary vector~$\bra{1}$, one down-spin is created for $s=0$
and the number of down-spins is kept invariant for $s=1$.

The tensor-network representation for the Bethe-state with open boundary conditions is shown in Fig.~\ref{fig:betheansatzopenbc}.
It contains twice as many rows as the tensor-network for periodic boundary conditions,
which makes the contraction more challenging, in principle.
However, as we see numerically, after a multiplication with a MPO-pair~$\mathcal{B}(\mu)$,
the Schmidt-rank of the state only increases by a factor of~$2$ - not $4$, as expected.
This suggests that there should exist a representation with virtual dimension~$2$ also in the open boundary conditions-case.


\section{The algebraic Bethe Ansatz} \label{sec:bethe}

Even though there exist numerous excellent reviews about the Algebraic Bethe
Ansatz~\cite{korepin93,gomez96,essler05,zvyagin05,plunkett09},
we resketch here the Ansatz in the picturesque Tensor-Network language
for sake of completeness. In this way, 
it is traceable, how the tensor networks shown in
Figs.~\ref{fig:betheansatz} and~\ref{fig:betheansatzopenbc} form exact eigenstates of integrable systems.


\subsection{The Yang-Baxter Algebra}

\begin{figure}[t]
    \begin{center}
        \includegraphics[width=0.48\textwidth]{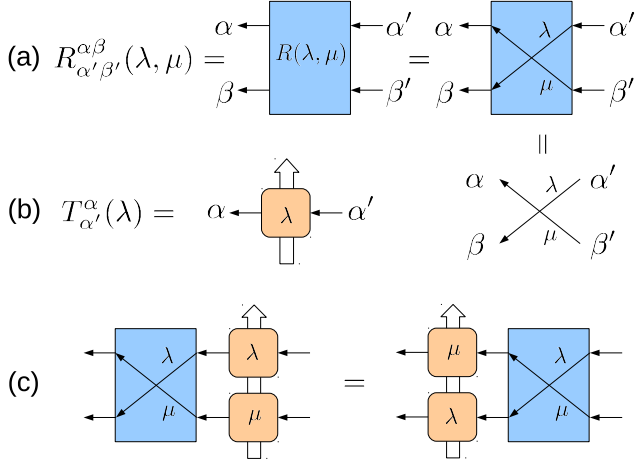}
     \end{center}
    \caption{
	(a) Visualization of the $4$-index $R$-tensor $R(\lambda,\mu)$.
	In the abbreviated version, it is visualized as two crossing arrows with $\lambda$
	attached to the up-down arrow and $\mu$ attached to the
	down-up arrow.
	(b) Yang-Baxter algebra as $4$-index tensor with two virtual indices (left-right)
	and two physical indices (up-down).
	(c) Defining equation for the Yang-Baxter algebra.
       }
    \label{fig:Rtensor}
\end{figure}

In general, the starting point for the Algebraic Bethe Ansatz is the $R(\lambda,\mu)$-tensor
\begin{equation} \label{eqn:R}
R^{\alpha \beta}_{\alpha' \beta'}(\lambda,\mu),
\end{equation}
with $\alpha,\beta,\alpha',\beta'$ ranging from $1$ to some ``auxiliary'' dimension~$d$
and $\lambda$, $\mu$ being some complex parameters.
This tensor defines the model under study, as will be shown later.
Graphically, the tensor is represented by two crossing arrows,
as shown in Fig.~\ref{fig:Rtensor}a,
where $\lambda$ and $\mu$ are associated to the up-down and down-up
arrows, respectively.
After joining indices $(\alpha \beta)$ and $(\alpha' \beta')$,
the tensor (\ref{eqn:R}) can also be interpreted as matrix $R(\lambda,\mu)$
acting on the vector space $V \otimes V$ (with $V=\mathbb{C}^d$).

\begin{figure}[t]
    \begin{center}
        \includegraphics[width=0.48\textwidth]{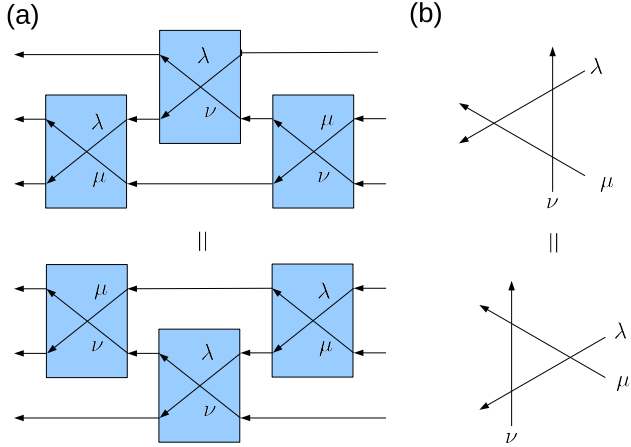}
     \end{center}
    \caption{
	(a) Yang-Baxter equation in tensor-network form.
	(b) Abbreviated version.
       }
    \label{fig:yangbaxter}
\end{figure}

\begin{figure}[t]
    \begin{center}
        \includegraphics[width=0.48\textwidth]{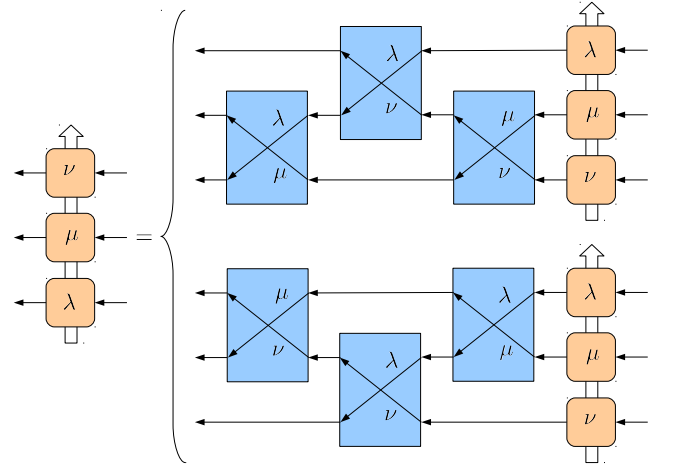}
     \end{center}
    \caption{
	Inversion of the ordering of 3 composed Yang-Baxter algebras using $R$-tensors.
	The inversion can be achieved in two ways, which makes
	necessary that the $R$-tensors fulfill the Yang-Baxter equation (Fig.~\ref{fig:yangbaxter}).
       }
    \label{fig:yangbaxtercondition}
\end{figure}

The condition on the $R$-tensor (\ref{eqn:R}) is 
that it fulfills Yang-Baxter equation (star-triangle relation).
Writing
\begin{eqnarray*}
R^{(23)} & = & \mathbb{1} \otimes R \\
R^{(12)} & = & R \otimes \mathbb{1},
\end{eqnarray*}
the Yang-Baxter equation reads
\begin{eqnarray*}
& R^{(23)} & (\lambda,\mu) R^{(12)}(\lambda,\nu) R^{(23)}(\mu,\nu) \\
& = & R^{(12)}(\mu,\nu) R^{(23)}(\lambda,\nu) R^{(12)}(\lambda,\mu).
\end{eqnarray*}
The graphical representation of this equation is shown in Fig.~\ref{fig:yangbaxtercondition}.
Another requirement is that solutions of the Yang-Baxter equation are regular,
meaning that there exists a $\lambda_0$ and a $\nu_0$, such that
\begin{equation} \label{eqn:regularity}
R^{\alpha \beta}_{\alpha' \beta'}(\lambda_0,\nu_0) = \delta^{\alpha}_{\alpha'} \delta^{\beta}_{\beta'}.
\end{equation}

The tensor $R(\lambda,\mu)$ defines the Yang-Baxter algebra
$T^{\alpha}_{\alpha'}(\lambda)$ ($\alpha,\alpha'=1,\ldots,d$)
by the relation
\begin{displaymath}
R^{\alpha \beta}_{\alpha' \beta'}(\lambda,\mu) T^{\alpha'}_{\alpha''}(\lambda) T^{\beta'}_{\beta''}(\mu) = 
T^{\alpha}_{\alpha'}(\mu) T^{\beta}_{\beta'}(\lambda) R^{\alpha' \beta'}_{\alpha'' \beta''}(\lambda,\mu)
\end{displaymath}
As usual, common indices are summed over.
Defining the Monodromy $T(\lambda)$ as the matrix of operators
\begin{displaymath}
T(\lambda) = 
\left(
\begin{array}{ccc}
T^1_1(\lambda) & \cdots & T^1_d(\lambda) \\
\vdots & \ddots & \vdots \\
T^d_1(\lambda) & \cdots & T^d_d(\lambda) \\
\end{array}
\right),
\end{displaymath}
the definition of the Yang-Baxter algebra can be written as
\begin{equation} \label{eqn:Ralgebra}
R(\lambda,\mu) \left[ T(\lambda) \check{\otimes} T(\mu) \right] = \left[ T(\mu) \check{\otimes} T(\lambda) \right] R(\lambda,\mu),
\end{equation}
where the outer product~``$\check{\otimes}$'' acts in the space $V \otimes V$ in the sense that
$\left[ T(\mu) \check{\otimes} T(\lambda) \right]^{\alpha \beta}_{\alpha' \beta'} \equiv T^{\alpha}_{\alpha'}(\mu) T^{\beta}_{\beta'}(\lambda)$.
$T(\lambda)$ can be considered as a $4$-index tensor: $2$ ``virtual'' indices $\alpha$, $\alpha'$ of dimension~$d$ select the
operator $T^{\alpha}_{\alpha'}(\mu)$ within the matrix, and two ``physical'' indices operate as input- and output
index of the operator. 
$T(\lambda)$ is represented graphically in Fig.~\ref{fig:Rtensor}b.
The virtual indices are indicated as horizontal arrows;
the physical input- and output indices are indicated as vertical in- and outgoing double-arrows.
Using this graphical notation, the definition of the Yang-Baxter algebra assumes the simple
form shown in Fig.~\ref{fig:Rtensor}c.

In this picture, $R(\lambda,\mu)$ has the property to permute the thensors $T(\lambda)$ and $T(\mu)$.
There is, however, one ambiguity that arises: there are two ways to go from
$T(\lambda) \check{\otimes} T(\mu) \check{\otimes} T(\nu)$ to 
$T(\nu) \check{\otimes} T(\mu) \check{\otimes} T(\lambda)$.
This inversion of the ordering can be achieved either
by exchanging firstly $\lambda \leftrightarrow \mu$, secondly $\lambda \leftrightarrow \nu$ and thirdly $\mu \leftrightarrow \nu$,
or by exchanging firstly $\nu \leftrightarrow \mu$, secondly $\lambda \leftrightarrow \nu$ and thirdly $\lambda \leftrightarrow \mu$.
This situation is depicted in Fig.~\ref{fig:yangbaxtercondition}.
Thus, both
\begin{eqnarray*}
& R^{(12)} & (\mu,\nu) R^{(23)}(\lambda,\nu) R^{(12)}(\lambda,\mu) \left[ T(\lambda) \check{\otimes} T(\mu) \check{\otimes} T(\nu) \right] \\
& = & \left[ T(\lambda) \check{\otimes} T(\mu) \check{\otimes} T(\nu) \right] R^{(12)}(\mu,\nu) R^{(23)}(\lambda,\nu) R^{(12)}(\lambda,\mu)
\end{eqnarray*}
and
\begin{eqnarray*}
& R^{(23)} & (\lambda,\mu) R^{(12)}(\lambda,\nu) R^{(23)}(\nu,\mu) \left[ T(\lambda) \check{\otimes} T(\mu) \check{\otimes} T(\nu) \right] \\
& = & \left[ T(\lambda) \check{\otimes} T(\mu) \check{\otimes} T(\nu) \right] R^{(23)}(\lambda,\mu) R^{(12)}(\lambda,\nu) R^{(23)}(\nu,\mu) 
\end{eqnarray*}
must be fulfilled.
These two equations, however, are compatible, because $R(\lambda,\mu)$ was required to fulfill the Yang-Baxter equation.
This makes the definition of the algebra~$T^{\alpha}_{\alpha'}(\lambda)$ consistent.

\begin{figure}[t]
    \begin{center}
        \includegraphics[width=0.48\textwidth]{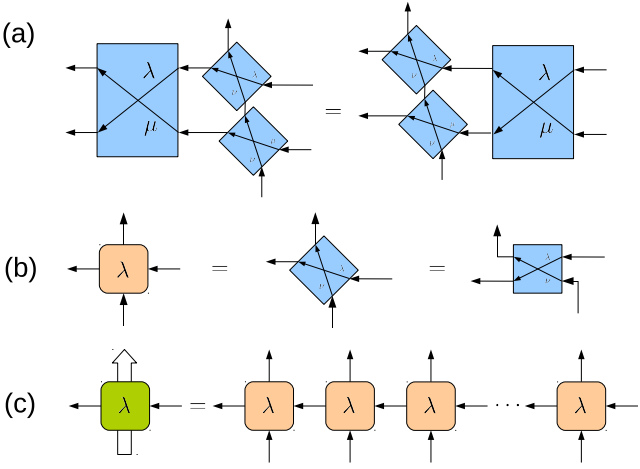}
     \end{center}
    \caption{
	(a) Deformation of the Yang-Baxter equation, such that it yields the
	fundamental representation of the Yang-Baxter algebra.
	(b) Connection between the fundamental representation and the $R$-tensor.
	(c) Formation of a more complex representation out of several
	fundamental representations.
       }
    \label{fig:fundamentalrepresentation}
\end{figure}

One representation of the Yang-Baxter algebra is easy to obtain - which is the
fundamental representation. 
This representation is formed by the operators $L^{\alpha}_{\alpha'}(\lambda,\nu)$
acting on $\mathbb{C}^d$ defined as
\begin{equation} \label{eqn:defL}
[ L^{\alpha}_{\alpha'}(\lambda,\nu) ]^k_l = R^{k \alpha}_{\alpha' l}(\lambda,\nu).
\end{equation}
In the graphical picture, the operators correspond to a clockwise ``rotation''
of the $R$-tensor by $45$ degrees, as shown in Fig.~\ref{fig:fundamentalrepresentation}b.
The two indices attached to the horizontal arrow then become the virtual indices of the operator,
and the vertical arrow carries the physical indices.
That these operators are a valid representation is due to the fact that the defining equation
\begin{displaymath}
R(\lambda,\mu) \left[ L(\lambda,\nu) \check{\otimes} L(\mu,\nu) \right] = 
\left[ L(\mu,\nu) \check{\otimes} L(\lambda,\nu) \right] R(\lambda,\mu)
\end{displaymath}
is just a ``distortion'' of the Yang-Baxter equation,
as shown in Fig.~\ref{fig:fundamentalrepresentation}a.
Up to now, the parameter $\nu$ in $L(\lambda,\nu)$ is arbitrary. Most conveniently it is to set $\nu = \nu_0$.

\begin{figure}[t]
    \begin{center}
        \includegraphics[width=0.48\textwidth]{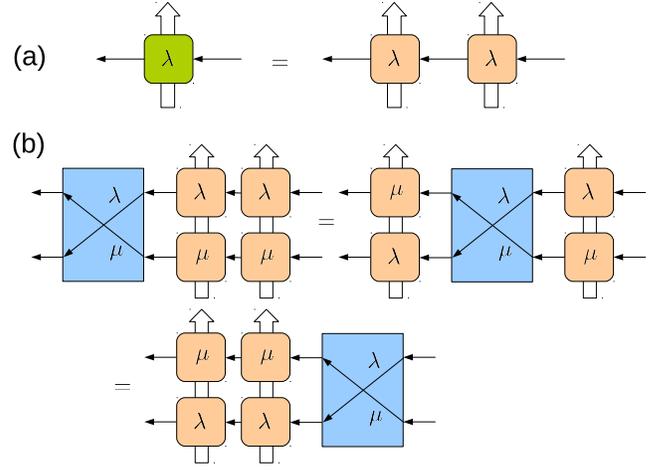}
     \end{center}
    \caption{
	(a) Co-multiplication property: formation of a new representation
	of the Yang-Baxter algebra out of two known representations.
	(b) Proof that the new representation still fulfills the
	defining equations for the Yang-Baxter algebra.
       }
    \label{fig:comultiplication}
\end{figure}

Once one representation $L(\lambda)$ is known, more complex representations are obtained
by concatenating the $L(\lambda)$'s horizontally, as depicted in Fig.~\ref{fig:fundamentalrepresentation}c.
Here, operators $T^{\alpha}_{\alpha'}(\lambda)$ acting on $(\mathbb{C}^d)^{\otimes N}$
are constructed out of $N$ simple operators $L^{\alpha}_{\alpha'}(\lambda)$ acting on $\mathbb{C}^d$ via
\begin{displaymath}
T^{\alpha}_{\alpha'}(\lambda) = \sum_{\alpha_2,\ldots,\alpha_N}
L^{\alpha}_{\alpha_2}(\lambda) \otimes
L^{\alpha_2}_{\alpha_3}(\lambda) \otimes
\cdots
\otimes
L^{\alpha_N}_{\alpha'}(\lambda).
\end{displaymath}
The outer product ``$\otimes$'' affects the physical indices.
In index notation, the operators read
\begin{displaymath}
[T^{\alpha}_{\alpha'}(\lambda)]^{k_1\cdots k_N}_{l_1\cdots l_N}
= \sum_{\alpha_2 \cdots \alpha_N}
[ L^{\alpha}_{\alpha_2}(\lambda) ]^{k_1}_{l_1}
[ L^{\alpha_2}_{\alpha_3}(\lambda) ]^{k_2}_{l_2}
\cdots
[ L^{\alpha_N}_{\alpha'}(\lambda) ]^{k_N}_{l_N}.
\end{displaymath}
The operators defined in such a way fulfill~(\ref{eqn:Ralgebra}),
because the $R$-tensor subsequently interchanges the operators $L^{\alpha}_{\alpha'}(\lambda)$
from left to right -- as can be retraced from Fig.~\ref{fig:comultiplication}b for $N=2$.
Defining the matrices $\mathcal{L}^k_l(\lambda)$ as
$\bra{\alpha} \mathcal{L}^k_l(\lambda) \ket{\alpha'} := [ L^{\alpha}_{\alpha'}(\lambda) ]^{k}_{l}$,
the operators $T^{\alpha}_{\alpha'}(\lambda)$ assume the form of MPOs,
\begin{displaymath}
T^{\alpha}_{\alpha'}(\lambda)
=
\sum_{\begin{smallmatrix}k_1 \cdots k_N\\l_1 \cdots l_N\end{smallmatrix} }
\bra{\alpha}
\mathcal{L}^{k_1}_{l_1}(\lambda)
\cdots
\mathcal{L}^{k_N}_{l_N}(\lambda)
\ket{\alpha'}
o^{k_1}_{l_1} \otimes \cdots \otimes o^{k_N}_{l_N},
\end{displaymath}
with $o^k_l = \ket{k}\bra{l}$.

\begin{figure}[t]
    \begin{center}
        \includegraphics[width=0.48\textwidth]{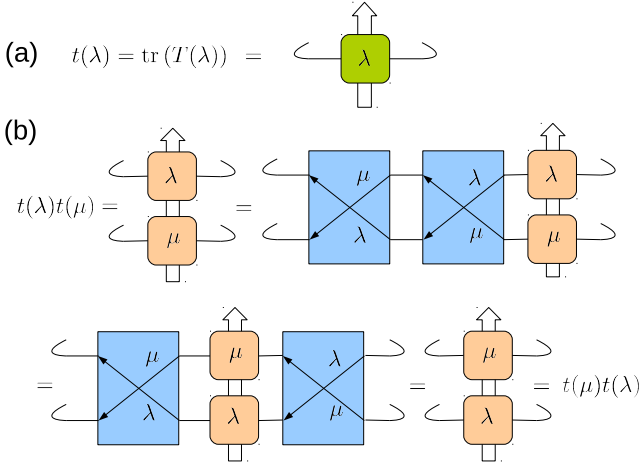}
     \end{center}
    \caption{
	(a) Definition of the transfer matrix.
	(b) Proof that all transfer matrices commute.
       }
    \label{fig:transfermatrix}
\end{figure}

The main building block of the Algebraic Bethe Ansatz is the transfer matrix $t(\lambda)$,
obtained as the trace of the algebra $T^{\alpha}_{\alpha'}(\lambda)$,
\begin{displaymath}
t(\lambda) := \tr \left\{ T(\lambda) \right\} \equiv \sum_{\alpha} T^{\alpha}_{\alpha}(\lambda).
\end{displaymath}
The transfer matrix $t(\lambda)$ corresponds to $T^{\alpha}_{\alpha'}(\lambda)$ with contracted
left and right indices $\alpha$ and $\alpha'$ (see Fig.~\ref{fig:transfermatrix}a).
In the MPO picture, $t(\lambda)$ is represented by an MPO with periodic boundary conditions.~\cite{verstraeteporras04}
Due to equation~($\ref{eqn:Ralgebra}$) that is fulfilled by the algebra, the transfer matrix has
the property that $[ t(\lambda), t(\mu) ] = 0$ for all $\lambda$ and $\mu$.
The way this property emerges from~($\ref{eqn:Ralgebra}$) can immediately be read off from Fig.~\ref{fig:transfermatrix}b:
starting out with the expression $t(\lambda) t(\mu)$,
the identity in the form $\mathbb{1} = R(\lambda,\mu)^{-1} R(\lambda,\mu)$ can be inserted
at the virtual bonds;
secondly, $R(\lambda,\mu)$ can be used to exchange $T(\lambda)$ and $T(\mu)$;
thirdly, the cyclic property of the trace can be used to elimiate $R(\lambda,\mu)$ and $R(\lambda,\mu)^{-1}$
in order to end up with $t(\mu) t(\lambda)$.

\begin{figure}[t]
    \begin{center}
        \includegraphics[width=0.48\textwidth]{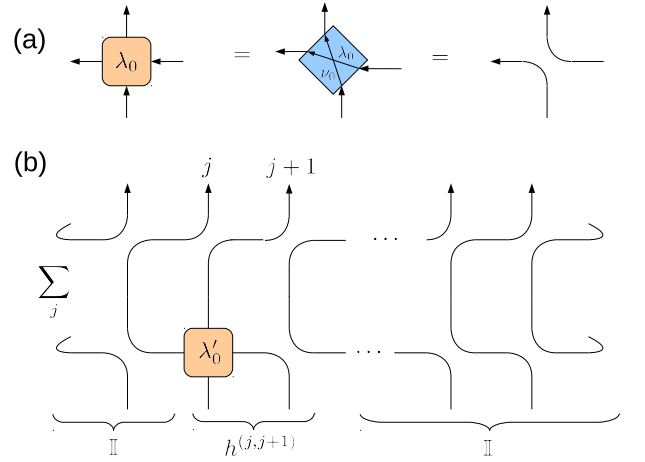}
     \end{center}
    \caption{
	(a) Precondition on the $R$-tensor: at some point $\lambda = \lambda_0$,
	the $R$-tensor decomposes into the outer product of two identities.
	(b) Logarithmic derivative of the transfer matrix $t(\lambda)$ at the
	point $\lambda=\lambda_0$. The first row represents $t(\lambda_0)^{-1}$,
	the second row $t'(\lambda_0)$.
       }
    \label{fig:hamiltonian}
\end{figure}

This property makes $t(\lambda)$ the generator of an infinite set of commuting observables:
if $t(\lambda)$ is Taylor-expanded with respect to $\lambda$,
$t(\lambda) = I_0 + \lambda I_1 + \lambda^2 I_2 + \ldots$, then
$[ I_j, I_k ] = 0$ for all $j$ and $k$. If one of the $I_k$'s is equal to the Hamiltonian of a model, it
is called \emph{integrable}, since there exist infinitely many symmetries which commute mutually.
In fact, any function of $t(\lambda)$ can be generating function for a set of commuting observables,
like e.g. $\mathcal{F}(\lambda)=\log t(\lambda)$.
The Taylor-expansion of this function reads
\begin{displaymath}
\mathcal{F}(\lambda) = \mathcal{F}(\lambda_0) + (\lambda-\lambda_0) \mathcal{F}'(\lambda_0) + O\left( (\lambda-\lambda_0)^2 \right).
\end{displaymath}
It turns out that $\mathcal{F}'(\lambda_0)$ is local in the sense that
\begin{displaymath}
\mathcal{F}'(\lambda_0) \equiv \frac{d}{d \lambda} \log{ t(\lambda) } \Big|_{\lambda=\lambda_0} = \sum_{i=1}^N h^{(i,i+1)}
\end{displaymath}
with $h^{(i,i+1)}$ only acting on sites $i$ and $i+1$.
Thus, an integrable model is obtained described by a local Hamiltonian
\begin{equation} \label{eqn:hamiltonian}
H = \sum_{i=1}^N h^{(i,i+1)}.
\end{equation}
Thereby,
\begin{equation} \label{eqn:defh}
h^{k_1 k_2}_{l_1 l_2} = \frac{d}{d \lambda} [ L^{k_1}_{l_2}(\lambda) ]^{k_2}_{l_1} \Big|_{\lambda=\lambda_0}
\end{equation}
or
\begin{displaymath}
h = \frac{\partial}{\partial \lambda} R(\lambda,\nu_0) \Big|_{\lambda=\lambda_0},
\end{displaymath}
respectively.
To see this connection, it has to be realized that due to the regularity condition~(\ref{eqn:regularity})
$t(\lambda_0)$ is equal to the cyclic shift operator that shifts the whole lattice to the right by one site.
The total momentum operator $\hat{P}$ is related to the cyclic shift operator according to
\begin{equation} \label{eqn:momentum}
e^{i \hat{P}} = t(\lambda_0).
\end{equation}
Graphically, $t(\lambda_0)$ is built from $L^{\alpha}_{\alpha'}(\lambda_0)$ shown in Fig.~\ref{fig:hamiltonian}a.
The way the local Hamiltonian~$H$ emerges by differentiating the non-local expression $\mathcal{F}(\lambda)$
is sketched in Fig.~\ref{fig:hamiltonian}b. Since $\mathcal{F}'(\lambda_0) = t(\lambda_0)^{-1} t'(\lambda_0)$,
the first row in the figure corresponds to the inverted cyclic shift operator $t(\lambda_0)^{-1}$
and the second row corresponds to the derivative $t'(\lambda_0)$. The derivative $t'(\lambda_0)$ disintegrates into
a sum of~$N$ derivatives with
respect to each of the tensors $L(\lambda)$ at sites $j=1,\ldots,N$.
As can be seen in the figure, term~$j$ has only support on two sites~$j$ and $j+1$
and thus corresponds to a two-site term that is related to the
derivative $L'(\lambda_0)$ as formulated in~(\ref{eqn:defh}).

Models that emerge in such a way from combinations of fundamental representations are fundamental models.
Examples are the spin-$1/2$ Heisenberg model and XXZ model.
In both cases, $d=2$ and the $R$-matrix assumes the form
\begin{equation} \label{eqn:Rgl2}
R(\lambda,\mu) = \left(
\begin{array}{cccc}
1 & & & \\
 & b(\lambda,\mu) & c(\lambda,\mu) & \\
 & c(\lambda,\mu) & b(\lambda,\mu) & \\
 & & & 1
\end{array}
\right)
\end{equation}
Also, $b(\lambda,\mu)$ and $c(\lambda,\mu)$ are of difference form, i.e.
$b(\lambda,\mu) = b(\lambda-\mu)$ and $c(\lambda,\mu) = c(\lambda-\mu)$.
This yields a $R$-matrix of difference form, as well: $R(\lambda,\mu)=R(\lambda-\mu)$.
Explicitly, the functions $b$ and $c$ read
\begin{eqnarray*}
b(\lambda) & = & \frac{1}{1+\lambda} \\
c(\lambda) & = & \frac{\lambda}{1+\lambda}
\end{eqnarray*}
for the Heisenberg model and
\begin{eqnarray}
b(\lambda) & = & \frac{\sinh(2 i \eta)}{\sinh(\lambda+2 i \eta)} \label{eqn:RXXZb}\\
c(\lambda) & = & \frac{\sinh(\lambda)}{\sinh(\lambda+2 i \eta)} \label{eqn:RXXZc}
\end{eqnarray}
for the XXZ model.
Evidently, in both cases, $R(0) = \mathbb{1}$, such that $\lambda_0 = 0$. 
In case of the Heisenberg model, $R'(0) = h_{XXX}$
with
\begin{displaymath}
h_{XXX} = \frac{1}{2} \left[ \sigma_x \otimes \sigma_x + \sigma_y \otimes \sigma_y + \sigma_z \otimes \sigma_z - \mathbb{1} \right].
\end{displaymath}
In the XXZ-case,
\begin{displaymath}
h_{XXZ}(\Delta) = \frac{1}{2} \left[ \sigma_x \otimes \sigma_x + \sigma_y \otimes \sigma_y + \Delta \left( \sigma_z \otimes \sigma_z - \mathbb{1} \right) \right]
\end{displaymath}
is obtained via
$R'(0) = 1/\sinh(2 i \eta) h_{XXZ}(\Delta)$ with $\Delta = \cos(2 \eta)$.

Models (fundamental and non-fundamental) with $R$-matrix~($\ref{eqn:Rgl2}$)
are $gl(2)$ generalized models. The Bethe ansatz for these models is especially simple
and will be described in the following.


\subsection{Bethe Ansatz for $gl(2)$ generalized models}

The Yang-Baxter Algebra with $R$-matrix~(\ref{eqn:Rgl2}) 
is generated by only $4$ elements, such that the Monodromy assumes the form
\begin{displaymath}
T(\lambda) = \left( \begin{array}{cc} A(\lambda) & B(\lambda)\\ C(\lambda) & D(\lambda) \end{array} \right)
\end{displaymath}
with
\begin{displaymath}
\begin{array}{ccclccc}
A(\lambda) & = & T^0_0(\lambda), & & C(\lambda) & = & T^1_0(\lambda)\\
B(\lambda) & = & T^0_1(\lambda), & & D(\lambda) & = & T^1_1(\lambda)
\end{array}.
\end{displaymath}
The most important commutation relations of the algebra are
\begin{eqnarray*}
B(\lambda) B(\mu) & = & B(\mu) B(\lambda) \\
A(\lambda) B(\mu) & = & \frac{1}{c(\mu,\lambda)} B(\mu) A(\lambda) - \frac{b(\mu,\lambda)}{c(\mu,\lambda)} B(\lambda) A(\mu)\\
D(\lambda) B(\mu) & = & \frac{1}{c(\lambda,\mu)} B(\mu) D(\lambda) - \frac{b(\lambda,\mu)}{c(\lambda,\mu)} B(\lambda) D(\mu).
\end{eqnarray*}
The precondition for the Ansatz is that a representation must exist, for which there is a
pseudo-vacuum~$\ket{vac}$ that is an eigenstate of
$A(\lambda)$ and $D(\lambda)$ and that is annihiliated by $C(\lambda)$:
\begin{eqnarray*}
A(\lambda) \ket{vac} & = & a(\lambda) \ket{vac}\\
D(\lambda) \ket{vac} & = & d(\lambda) \ket{vac}\\
C(\lambda) \ket{vac} & = & 0.
\end{eqnarray*}

The goal is to diagonalize the transfer matrix $t(\lambda) = A(\lambda) + D(\lambda)$.
Since all transfer matrices commute, $[t(\lambda),t(\mu)]=0$, they have a common system of eigenvectors.
Thus, all eigenvectors are independent of $\lambda$. The eigenvalue problem reads
\begin{displaymath}
t(\lambda) \ket{\Psi} = \tau(\lambda) \ket{\Psi}.
\end{displaymath}
The Bethe Ansatz
\begin{displaymath}
\ket{\Psi(\mu_1,\ldots,\mu_M)} = B(\mu_1) \cdots B(\mu_M) \ket{vac}.
\end{displaymath}
fulfills the eigenvalue problem provided that the $\mu_k$'s fulfill the
Bethe equations
\begin{equation} \label{eqn:bethe}
\frac{d(\mu_n)}{a(\mu_n)} =
\prod_{\begin{smallmatrix} j=1\\ j\neq n \end{smallmatrix}}^M
\frac{c(\mu_n,\mu_j)}{c(\mu_j,\mu_n)}
\end{equation}
(n=1,\ldots,M).
The eigenvalue $\tau(\lambda)$ is then equal to
\begin{displaymath}
\tau(\lambda) =
a(\lambda) \prod_{j=1}^M \frac{1}{c(\mu_j,\lambda)} +
d(\lambda) \prod_{j=1}^M \frac{1}{c(\lambda,\mu_j)}.
\end{displaymath}
The proof is obtained by utilizing algebraic relations only and can be gathered from appendix~\ref{app:derivbethe}.
From $\tau(\lambda)$, the eigenvalue of the
Hamiltonian~(\ref{eqn:hamiltonian}) is obtained as
\begin{equation} \label{eqn:energyev}
E = \frac{\tau'(\lambda_0)}{\tau(\lambda_0)}.
\end{equation}
The total momentum is, according to~(\ref{eqn:momentum}), equal to
\begin{equation} \label{eqn:momentumev}
p = -i \ln \tau(\lambda_0).
\end{equation}


\subsection{Bethe Ansatz for the Heisenberg model and the XXZ~model}

In case of the Heisenberg model and XXZ model, this spezializes as follows:
the matrices $\mathcal{L}^{k}_{l}(\lambda)$ that build up the MPOs $T^{\alpha}_{\alpha'}(\lambda)$
have block form.
Written out, they read
\begin{displaymath}
\begin{array}{cc}
\mathcal{L}^{0}_{0}(\lambda) = \left( \begin{array}{cc} 1 & 0 \\ 0 & c(\lambda) \end{array} \right), &
\mathcal{L}^{0}_{1}(\lambda) = \left( \begin{array}{cc} 0 & 0 \\ b(\lambda) & 0 \end{array} \right) \\ \\
\mathcal{L}^{1}_{0}(\lambda) = \left( \begin{array}{cc} 0 & b(\lambda) \\ 0 & 0 \end{array} \right), &
\mathcal{L}^{1}_{1}(\lambda) = \left( \begin{array}{cc} c(\lambda) & 0 \\ 0 & 1 \end{array} \right) 
\end{array}
\end{displaymath}
These MPOs are symmetry conserving in the sense that $T^{\alpha}_{\alpha'}(\lambda)$ changes the number of
down-spins by $\alpha'-\alpha$.
This is due to the local constraint that 
$[\mathcal{L}^{k}_{l}(\lambda)]^{\alpha}_{\alpha'}$ are non-zero only if
$\alpha' = \alpha + (k-l)$, as discussed in Sec.~\ref{sec:mpsandbethe}.
 
Using these considerations, the vacuum state is obviously the state with no down-spins, namely
\begin{displaymath}
\ket{vac} = \ket{0} \otimes \cdots \otimes \ket{0}
\end{displaymath}
($0 \equiv \uparrow$, $1 \equiv \downarrow$) .
This state is annihilated by $C(\lambda)$
and is an eigenvector of $A(\lambda)$ and $D(\lambda)$ with eigenvalues
\begin{displaymath}
a(\lambda)=1, \qquad d(\lambda)=c(\lambda)^N.
\end{displaymath}
The Bethe-Ansatz state
\begin{displaymath}
\ket{\Psi(\mu_1,\ldots,\mu_M)} = B(\mu_1) \cdots B(\mu_M) \ket{vac}
\end{displaymath}
is a state with $M$ down-spins, i.e. with magnetization in
$z$-direction equal to $S_z = \frac{1}{2} N - M$.

The Bethe equations obtained by the Algebraic Bethe Ansatz
are equal to the equations obtained by the Coordinate Bethe Ansatz.
In case of Heisenberg model, it is advantageous to
introduce variables $z_j$ that are related to $\mu_j$ in~(\ref{eqn:bethe}) via
\begin{displaymath}
\mu_j = \frac{z_j}{2i} - \frac{1}{2}
\end{displaymath}
for a direct comparison with results of coordinate Bethe ansatz~\cite{bethe31,hulthen38,karbach98a,karbach98b,karbach00}:
In terms of these variables, the Bethe equations read
\begin{equation} \label{eqn:betheheisenberg}
\left( \frac{z_n - i}{z_n + i} \right)^N =
\prod_{\begin{smallmatrix} j=1\\ j\neq n \end{smallmatrix}}^M
\frac{z_n - z_j - 2i}{z_n - z_j + 2i}
\end{equation}
with $n=1,\ldots,M$. From the Bethe solutions $\{ z_j \}$, the energy is obtained
using~(\ref{eqn:energyev}) as
\begin{displaymath}
E = \frac{\tau'(0)}{\tau(0)} = - \sum_{j=1}^M \frac{4}{z_j^2 + 1}.
\end{displaymath}
According to~(\ref{eqn:momentumev}), the total momentum~$\hat{P}$ has eigenvalue
\begin{displaymath}
p = -i \ln \tau(0) = \sum_{j=1}^M \left( -i \ln \frac{z_j+i}{z_j-i} \right).
\end{displaymath}
The addends are usually referred to as magnon momenta that can be written as
\begin{displaymath}
p_j = \pi - 2 \arctan(z_j).
\end{displaymath}
using the identity
\begin{displaymath}
\arctan(z) = \frac{1}{2 i} \ln \frac{1 + i z}{1 - i z}.
\end{displaymath}
In term of the magnon momenta~$p_j$, the total momentum reads
\begin{equation} \label{eqn:xxxmomentum}
p = \sum_{j=1}^M p_j
\end{equation}
and the energy is equal to
\begin{displaymath}
E = - 2 \sum_{j=1}^M \left( 1-\cos(p_j) \right).
\end{displaymath}
For solving the Bethe equations~(\ref{eqn:betheheisenberg}) it is advantageous to
bring them to their their logarithmic form
\begin{displaymath}
N p_n = 2 \pi I_n + \sum_{\begin{smallmatrix} j=1\\ j\neq n \end{smallmatrix}}^M \Theta(p_n,p_j),
\end{displaymath}
where
\begin{displaymath}
2 \cot \frac{\Theta(p,q)}{2} =
\cot \frac{p}{2} -
\cot \frac{q}{2}.
\end{displaymath}
and $I_j$ are integers $\in \{0,\ldots,N\}$.
Solutions can then be found iteratively, as described in~[\onlinecite{karbach98b}].
The ground state configuration for $N$ even and $M=N/2$ is $(I_1,\ldots,I_M) = (1,3,\ldots,N-1)$.

In case of XXZ model, it is advantageous to
introduce the variables $z_j$ related to $\mu_j$ in~(\ref{eqn:bethe}) via
\begin{displaymath}
\mu_j = z_j - i \eta + i \frac{\pi}{2}
\end{displaymath}
to compare with the Coordinate Bethe Ansatz~\cite{yangyang66,orbach58}.
The Bethe equations then read
\begin{displaymath}
\left( \frac{ \cosh(z_n - i \eta) }{ \cosh(z_n + i \eta } \right)^N =
\prod_{\begin{smallmatrix} j=1\\ j\neq n \end{smallmatrix}}^M
\frac{\sinh(z_n - z_j - 2 i \eta)}{\sinh(z_n - z_j + 2 i \eta)}.
\end{displaymath}
From the Bethe solutions $\{ z_j \}$, the energy is obtained as
\begin{displaymath}
E = \sinh(2 i \eta) \frac{\tau'(0)}{\tau(0)} = 
2 \sum_{j=1}^M \frac{\sin(2 \eta)^2}{\cos(2 \eta) + \cosh(2 z_j)}.
\end{displaymath}
The total momentum obtained from~(\ref{eqn:momentumev}) is
again of form~(\ref{eqn:xxxmomentum}) with
\begin{displaymath}
p_j = -2 \arctan \left( \tanh(z_j) \tan(\eta) \right).
\end{displaymath}
In terms of the momenta~$p_j$, the energy can be expressed as
\begin{displaymath}
E = - 2 \sum_{j=1}^M \left( \Delta-\cos(p_j) \right).
\end{displaymath}
The Bethe equations in their logarithmic form read
\begin{displaymath}
N p_n = 2 \pi I_n + \sum_{\begin{smallmatrix} j=1\\ j\neq n \end{smallmatrix}}^M \Theta(p_n,p_j),
\end{displaymath}
with
\begin{equation} \label{eqn:deftheta}
\cot \frac{\Theta(p,q)}{2} =
\frac{\Delta \sin \frac{p-q}{2} }{\cos \frac{p+q}{2} - \Delta \cos \frac{p-q}{2} }
\end{equation}
and $I_j$ $\in$ $\{0,\ldots,N\}$.
The ground state configuration for $N$ even and $M=N/2$
is again found with $(I_1,\ldots,I_M) = (1,3,\ldots,N-1)$.


\section{Algebraic Bethe Ansatz for open boundary conditions} \label{sec:openbc}

The method described for periodic boundary conditions is
generalizeable to models with open boundary conditions and boundary fields~\cite{cherednik84,sklyanin88,kitanine07}.
We resketch here the Ansatz for open boundary conditions
following closely Sklyanin~\cite{sklyanin88}
using a picturesque language.

\begin{figure}[t]
    \begin{center}
        \includegraphics[width=0.48\textwidth]{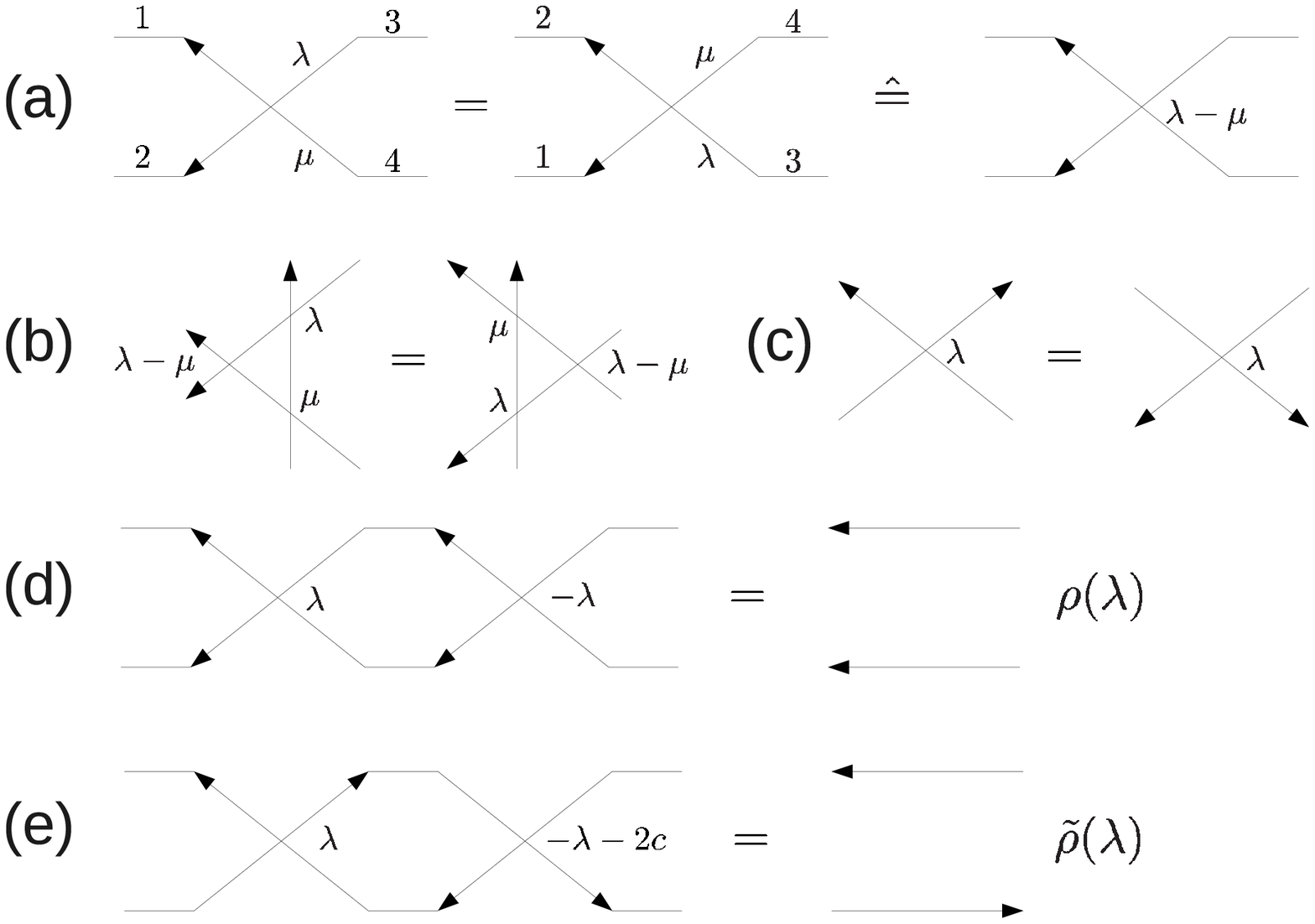}
     \end{center}
    \caption{
	(a) Permutation symmetry.
	(b) Yang-Baxter equation.
	(b) Partial transpostion symmetry.
	(e) Crossing unitary condition.
	(d) Unitary condition.
       }
    \label{fig:definitionsopenbc}
\end{figure}

For the following it is required that the $R$-tensor fulfills several conditions.
To express these, it is convenient to define the permutation operator
\begin{displaymath}
P = \sum_{i,j} \ket{j,i} \bra{i,j}
\end{displaymath}
that permutes two indices.
Using the matrix-notation $R(\lambda,\mu)$ from appendix~\ref{sec:bethe}, 
i.e. considering the $R$-tensor as matrix acting on $V \otimes V$ with $V = \mathbb{C}^d$,
a variant of the $R$-tensor with the first two indices permuted can be defined as
\begin{displaymath}
\mathcal{R}(\lambda,\mu) = P R(\lambda,\mu)
\end{displaymath}
The basic assumption is that the
$R$-tensor fullfills the symmetry condition
\begin{displaymath}
P R(\lambda,\mu) P = R(\lambda,\mu)
\end{displaymath}
(see Fig.~\ref{fig:definitionsopenbc}a).
Then, the $R$-tensor can expressed just by two crossing arrows and it is not necessary to
distinguish between them by marking them with the arguments.
In fact, it is assumed in the following that $R$ is of 
difference form, i.e. $R(\lambda,\mu) = R(\lambda-\mu)$.
Thus, the tensor $R(\lambda-\mu)$ will be characterized by
two crossing arrows together with the argument $\lambda-\mu$,
as shown by the rightmost depiction in Fig.~\ref{fig:definitionsopenbc}a.
Using this notation, the Yang-Baxter equation assumes the form shown in Fig.~\ref{fig:definitionsopenbc}b.

It is furthermore useful to define the partial transposition
\begin{displaymath}
[ \mathcal{R}(\lambda)^{t_1} ]^{\alpha \beta}_{\alpha' \beta'} = [ \mathcal{R}(\lambda) ]^{\alpha' \beta}_{\alpha \beta'},
\end{displaymath}
which is equivalent to flipping the direction of ``up-down'' arrow.
In analogy,
\begin{displaymath}
[ \mathcal{R}(\lambda)^{t_2} ]^{\alpha \beta}_{\alpha' \beta'} = [ \mathcal{R}(\lambda) ]^{\alpha \beta'}_{\alpha' \beta}
\end{displaymath}
corresponds to flipping the direction of the ``down-up'' arrow.
Accordingly, the partial transposition symmetry condition
\begin{displaymath}
\mathcal{R}(\lambda)^{t_1} = \mathcal{R}(\lambda)^{t_2}
\end{displaymath}
is expressed by Fig.~\ref{fig:definitionsopenbc}c.

Further conditions are the unitarity condition
\begin{equation} \label{eqn:unitarityopenbc}
\mathcal{R}(\lambda) \mathcal{R}(-\lambda) = \rho(\lambda)
\end{equation}
and the crossing unitarity condition
\begin{equation} \label{eqn:crossingunitarityopenbc}
\mathcal{R}(\lambda)^{t_1} \mathcal{R}(-\lambda-2 c)^{t_1} = \tilde{\rho}(\lambda)
\end{equation}
with $\rho(\lambda)$ and $\tilde{\rho}(\lambda)$ being some scalar functions of $\lambda$
and $c$ denoting some constant characterizing the $R$-tensor.
These conditions are represented by Figs.~\ref{fig:definitionsopenbc}d and~\ref{fig:definitionsopenbc}e.

\subsection{Reflection Algebras}

\begin{figure}[t]
    \begin{center}
        \includegraphics[width=0.48\textwidth]{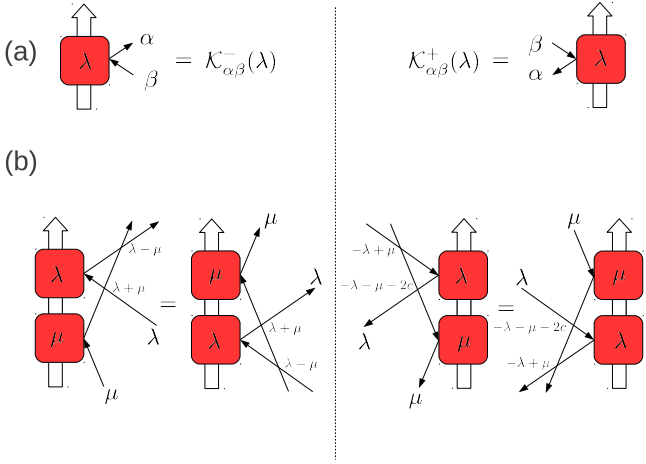}
     \end{center}
    \caption{
	(a) Graphical representation of the reflection algebras $\mathcal{K}^-(\lambda)$ and $\mathcal{K}^+(\lambda)$.
	The horizontal arrows indicate the virtual indices $\alpha$ and $\beta$;
	the vertical arrow indicates physical indices, i.e. the input- and output indices of the operators
	$\mathcal{K}^-_{\alpha \beta}(\lambda)$ and $\mathcal{K}^+_{\alpha \beta}(\lambda)$ respectively.
	(b) Defining equations for the reflection algebras (reflection equations).
	Each intersection of two lines represents an $R$-tensor. The argument of the $R$-tensor
	is written next to the intersection.
       }
    \label{fig:reflectionequationsopenbc}
\end{figure}

As the Bethe Ansatz for periodic boundary conditions is based on the Yang-Baxter algebra, 
the footing of the open boundary conditions Ansatz are the reflection algebras
$\mathcal{K}^-(\lambda)$ and $\mathcal{K}^+(\lambda)$
spanned by $\{ \mathcal{K}^-_{\alpha \beta}(\lambda) | \alpha,\beta=1,\ldots,d\}$
and $\{ \mathcal{K}^+(\lambda)_{\alpha \beta} | \alpha,\beta=1,\ldots,d\}$.
The graphical representation of these two algebras is shown in Fig.~\ref{fig:reflectionequationsopenbc}a:
as in the case of the Yang-Baxter algebra,
each of the two algebras is considered as a $4$-index tensor with
$2$ ``virtual'' indices $\alpha$ and $\beta$ of dimension~$d$, represented by the horizontal arrows,
and the $2$ ``physical'' indices
(corresponding to the input- and output indices of the operators
$\mathcal{K}^-_{\alpha \beta}(\lambda)$ and $[\mathcal{K}^+_{\alpha \beta}(\lambda)]$ respectively),
represented by the vertical arrows.
The only difference to the Yang-Baxter case is that the virtual indices both are on the right-hand side
of the tensor in case of $\mathcal{K}^-(\lambda)$ and on the left-hand side in case of $\mathcal{K}^+(\lambda)$.
The correspondence to the Monodromy in the open boundary condition case is
the matrix of operators
\begin{displaymath}
\mathcal{K}^{\pm}(\lambda) = 
\left(
\begin{array}{ccc}
\mathcal{K}^{\pm}_{1 1}(\lambda) & \cdots & \mathcal{K}^{\pm}_{1 d}(\lambda) \\\relax
\vdots & \ddots & \vdots \\\relax
\mathcal{K}^{\pm}_{d 1}(\lambda) & \cdots & \mathcal{K}^{\pm}_{d d}(\lambda)
\end{array}
\right).
\end{displaymath}

The defining equations for the reflection algebras are the reflection equations, represented by 
the tensor network in Fig.~\ref{fig:reflectionequationsopenbc}b.
In this figure, each intersection of two lines represents an $R$-tensor. The argument of the $R$-tensor
is written next to the intersection.
Algebraically, the reflection equations read
\begin{eqnarray*}
& \mathcal{R}(\lambda-\mu)
\overset{1}{\mathcal{K}} \vphantom{10pt}^{-}(\lambda)
\mathcal{R}(\lambda+\mu)
\overset{2}{\mathcal{K}} \vphantom{10pt}^{-}(\mu) \hspace{5cm} &\\
& =
\overset{2}{\mathcal{K}} \vphantom{10pt}^{-}(\mu)
\mathcal{R}(\lambda+\mu)
\overset{1}{\mathcal{K}} \vphantom{10pt}^{-}(\lambda)
\mathcal{R}(\lambda-\mu) &\\
\end{eqnarray*}
and
\begin{eqnarray*}
& \mathcal{R}(-\lambda+\mu)
[\overset{1}{\mathcal{K}} \vphantom{10pt}^{+}(\lambda)]^{t_1}
\mathcal{R}(-\lambda-\mu-2 c)
[\overset{2}{\mathcal{K}} \vphantom{10pt}^{+}(\mu)]^{t_2} \hspace{2cm} &\\
& =
[\overset{2}{\mathcal{K}} \vphantom{10pt}^{+}(\mu)]^{t_2}
\mathcal{R}(-\lambda-\mu-2 c)
[\overset{1}{\mathcal{K}} \vphantom{10pt}^{+}(\lambda)]^{t_1}
\mathcal{R}(-\lambda+\mu). &\\
\end{eqnarray*}
with
\begin{eqnarray*}
\overset{1}{\mathcal{K}} \vphantom{10pt}^{\pm}(\lambda) & = & \mathcal{K}^{\pm}(\lambda) \check{\otimes} \mathbb{1} \\
\overset{2}{\mathcal{K}} \vphantom{10pt}^{\pm}(\lambda) & = & \mathbb{1} \check{\otimes} \mathcal{K}^{\pm}(\lambda).
\end{eqnarray*}
The outer product~``$\check{\otimes}$'' is thereby interpreted as in~(\ref{eqn:Ralgebra})
and $\mathbb{1}$ is the $d \times d$ identity matrix.

\begin{figure}[t]
    \begin{center}
        \includegraphics[width=0.48\textwidth]{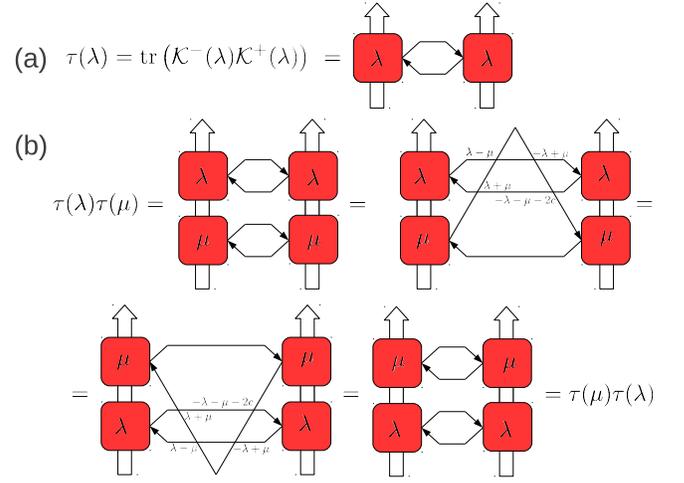}
     \end{center}
    \caption{
	(a) Definition of the transfer matrix.
	(b) Proof of the commuting property of the transfer matrix, $[\tau(\lambda),\tau(\mu)]=0$.
       }
    \label{fig:transfermatrixpopenbc}
\end{figure}

Using these algebras, it is possible to define a commuting set of transfer matrices via
\begin{displaymath}
\tau(\lambda) = \mathrm{tr} \left( \mathcal{K}^-(\lambda) \mathcal{K}^+(\lambda) \right).
\end{displaymath}
Graphically, $\tau(\lambda)$ corresponds to $\mathcal{K}^-(\lambda)$ and $\mathcal{K}^+(\lambda)$ being
glued together, as shown in Fig.~\ref{fig:transfermatrixpopenbc}a.
The commutativity of the transfer matrices, $[\tau(\lambda),\tau(\mu)]=0$, can
be proven using the unitary and crossing unitary conditions~(\ref{eqn:unitarityopenbc}) and (\ref{eqn:crossingunitarityopenbc})
and the reflection equations. The proof is sketched in Fig.~\ref{fig:transfermatrixpopenbc}b:
starting out with $\tau(\lambda) \tau(\mu)$, the line connecting $\mathcal{K}^+(\mu)$ and $\mathcal{K}^-(\mu)$
can be pulled over the line lying above that connects $\mathcal{K}^-(\lambda)$ and $\mathcal{K}^+(\lambda)$ using~(\ref{eqn:crossingunitarityopenbc})
and over the topmost arrow connecting the two $\lambda$-algebras using~(\ref{eqn:unitarityopenbc}).
Next, the network is mirrored vertically by using the reflection equations.
Finally, the drawn out line is pushed back using~(\ref{eqn:unitarityopenbc}) and (\ref{eqn:crossingunitarityopenbc}),
which leads to $\tau(\mu) \tau(\lambda)$, as desired.

\begin{figure}[t]
    \begin{center}
        \includegraphics[width=0.48\textwidth]{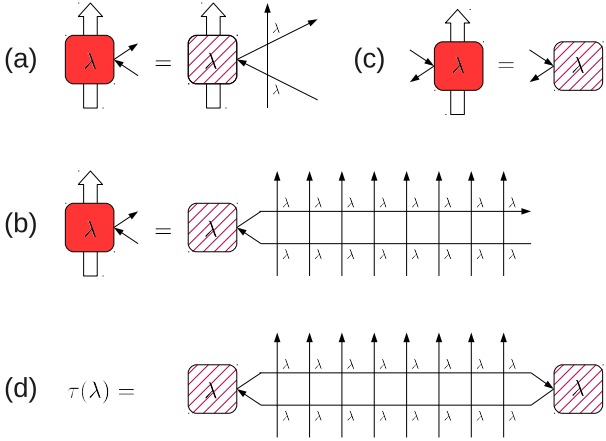}
     \end{center}
    \caption{
	(a) Composition of a new representation of $\mathcal{K}^-(\lambda)$ out of one pair of $R$-tensors and
	a known representation of $\mathcal{K}^-(\lambda)$ (that already fulfills the reflection equations).
	The known representation is indicated by the shaded surface.
	(b) Simple representation of $\mathcal{K}^+(\lambda)$ with physical dimension~$1$.
	(c) Composition of a complex representations of $\mathcal{K}^-(\lambda)$ of dimension~$d^N$ by
	attaching $N$ pairs of $R$-tensors to a simple representation with physical dimension~$1$.
	(d) Transfer matrix built from the representations (b) and (c).
       }
    \label{fig:fundamentalrepopenbc}
\end{figure}

\begin{figure}[t]
    \begin{center}
        \includegraphics[width=0.48\textwidth]{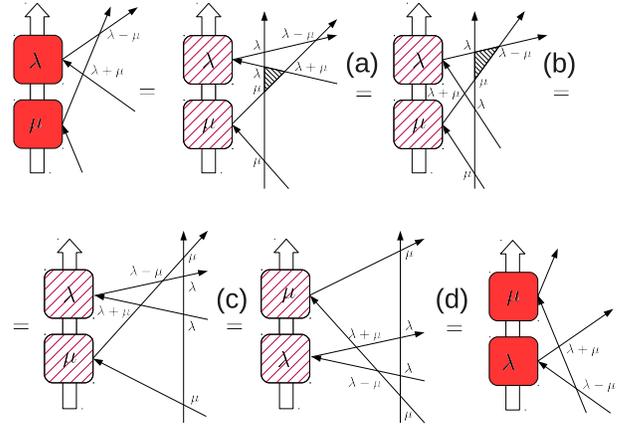}
     \end{center}
    \caption{
	Proof that the composed representation shown in Fig.~\ref{fig:fundamentalrepopenbc}a
	fulfills the reflection equations:
	the main idea of the first two steps (a) and (b) is to pull the vertical line rightmost by applying the
	Yang-Baxter equation twice. The three $R$-tensors to which the Yang-Baxter equation is applied
	are marked by the shaded triangles.
	The new situation now allows the application of the reflection equations, as shown in step (c).
	The last step (d) consists in pushing the vertical line back by applying the Yang-Baxter equation twice,
	such as in steps (a) and (b), but in reverse order.
       }
    \label{fig:prooffundamentalrepopenbc1}
\end{figure}

What is remaining is to find concrete representations of the reflection algebras.
Examples of simple representations with physical dimension~$1$ have already been found.~\cite{cherednik84}
More complex representations can be constructed by assembling a known representation with two $R$-tensors
in the way shown in Fig.~\ref{fig:fundamentalrepopenbc}a.
The physical dimension of the new representation is thereby increased by a factor~$d$.
That this assembly is indeed a valid representation can be proven
using the Yang-Baxter equation and the reflection equations.
The proof is sketched in Fig.~\ref{fig:prooffundamentalrepopenbc1}.

Thus, starting out with a simple representation with physical dimension~$1$ for $\mathcal{K}^-(\lambda)$,
a representation with physical dimension~$d^N$ is obtained after $N$ iterations 
with the relation expressed in Fig.~\ref{fig:fundamentalrepopenbc}a.
The structure of the representation after $N$ iterations can be gathered from Fig.~\ref{fig:fundamentalrepopenbc}b.
Assuming a simple representation with physical dimension~$1$ for $\mathcal{K}^+(\lambda)$
(depicted in Fig.~\ref{fig:fundamentalrepopenbc}c),
the transfer matrix assumes the form shown in Fig.~\ref{fig:fundamentalrepopenbc}d.

\begin{figure}[t]
    \begin{center}
        \includegraphics[width=0.48\textwidth]{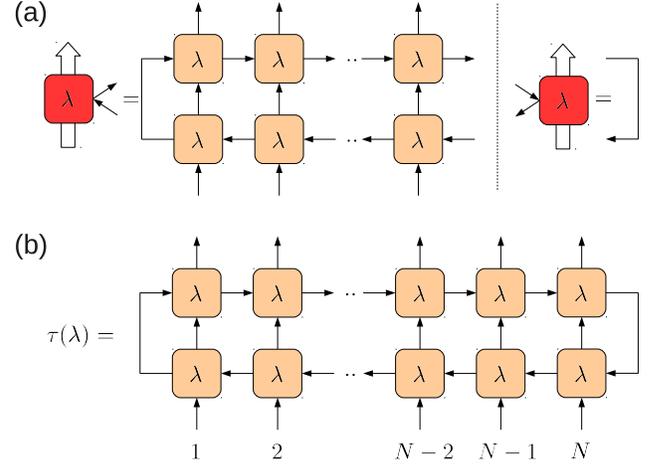}
     \end{center}
    \caption{
	(a) Representation of $\mathcal{K}^-(\lambda)$ and $\mathcal{K}^+(\lambda)$ for open boundary conditions.
	(b) Transfer matrix built from these representations.
       }
    \label{fig:fundamentalrepopenbc2}
\end{figure}

For the sake of simplicity, we choose the simple representations with physical dimension~$1$ equal to the identity
(which is a valid representation that fulfills the reflection equations).
Using the notation for the fundamental representation of the Yang-Baxter algebra introduced in 
equation~(\ref{eqn:defL}) and Fig.~\ref{fig:fundamentalrepresentation}b,
the representations of the algebras $\mathcal{K}^-(\lambda)$ and $\mathcal{K}^+(\lambda)$ 
look as shown in Fig.~\ref{fig:fundamentalrepopenbc2}a.
The transfer matrix assumes the form depicted in Fig.~\ref{fig:fundamentalrepopenbc2}b.
Algebraically, the representation of $\mathcal{K}^-(\lambda)$ is then
the product of two MPOs,
\begin{equation} \label{eqn:krepopenbc}
\mathcal{K}^-_{\alpha \beta}(\lambda) = \sum_{s=1}^d \bar{K}^-_{s \alpha}(\mu) K^-_{s \beta}(\mu).
\end{equation}
In terms of the previously defined matrices $\mathcal{L}^{k}_{l}(\mu)$, the MPOs read
\begin{displaymath}
K^-_{s \beta}(\mu)
=
\sum_{\begin{smallmatrix}k_1 \cdots k_N\\l_1 \cdots l_N\end{smallmatrix} }
\bra{s}
\mathcal{L}^{k_1}_{l_1}(\mu)
\cdots
\mathcal{L}^{k_N}_{l_N}(\mu)
\ket{\beta}
o^{k_1}_{l_1} \otimes \cdots \otimes o^{k_N}_{l_N}
\end{displaymath}
and
\begin{displaymath}
\bar{K}^-_{s \alpha}(\mu)
=
\sum_{\begin{smallmatrix}k_1 \cdots k_N\\l_1 \cdots l_N\end{smallmatrix} }
\bra{s}
\mathcal{L}^{k_1}_{l_1}(\mu)^T
\cdots
\mathcal{L}^{k_N}_{l_N}(\mu)^T
\ket{\alpha}
o^{k_1}_{l_1} \otimes \cdots \otimes o^{k_N}_{l_N}
\end{displaymath}
with $o^k_l = \ket{k} \bra{l}$.
The representation of $\mathcal{K}^+(\lambda)$ 
has physical dimension one and is equal to the identity with respect to the virtual indices, i.e.
\begin{displaymath}
\mathcal{K}^+_{\alpha \beta}(\lambda) = \delta_{\alpha \beta}.
\end{displaymath}

\begin{figure}[t]
    \begin{center}
        \includegraphics[width=0.48\textwidth]{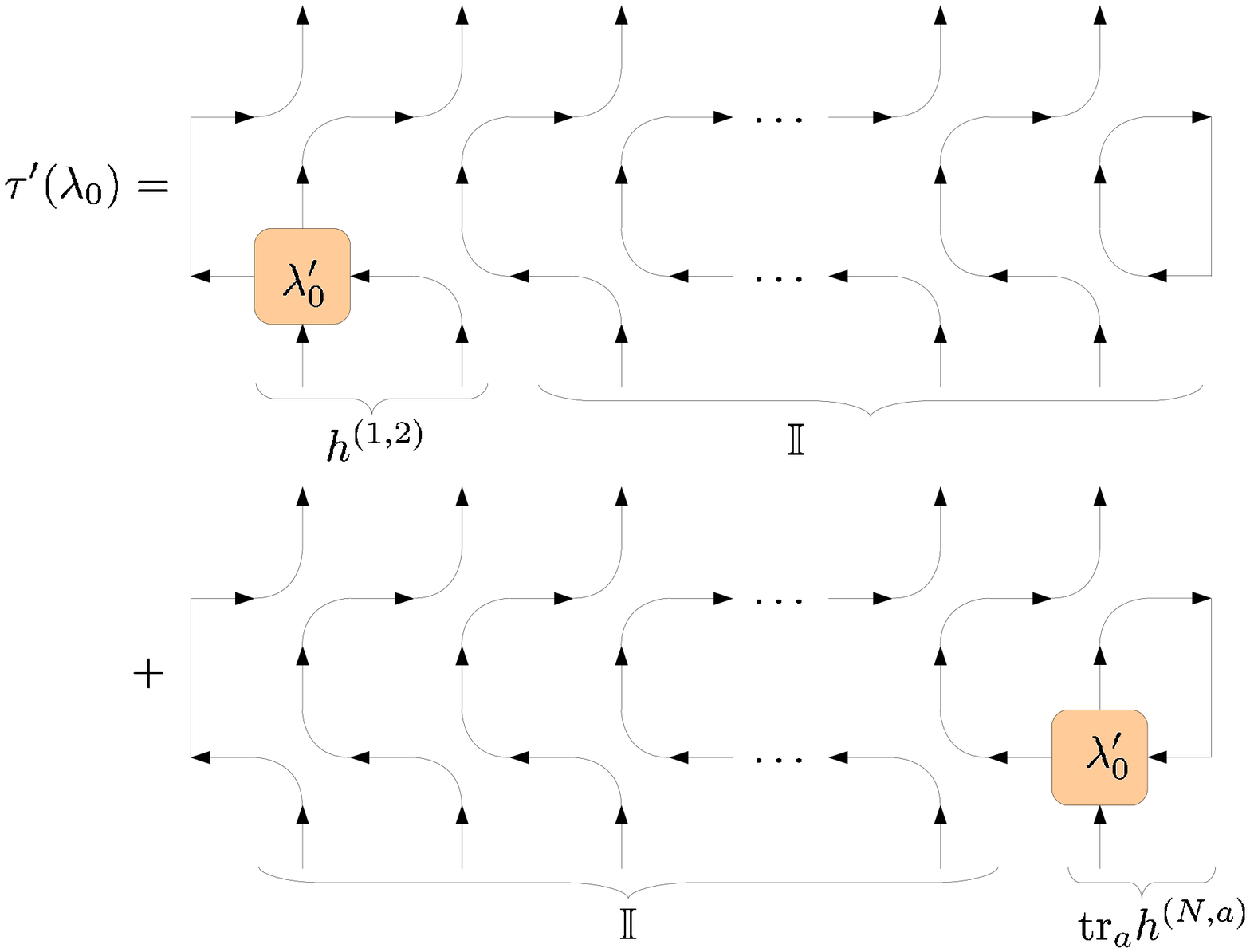}
     \end{center}
    \caption{
	Derivation of the open boundary condition Hamiltonian by
	derivative of the transfer matrix shown in Fig.~\ref{fig:fundamentalrepopenbc2}b
	at the point $\lambda_0$.
	(part~I).
       }
    \label{fig:hamiltonianopenbc}
\end{figure}

\begin{figure}[t]
    \begin{center}
        \includegraphics[width=0.48\textwidth]{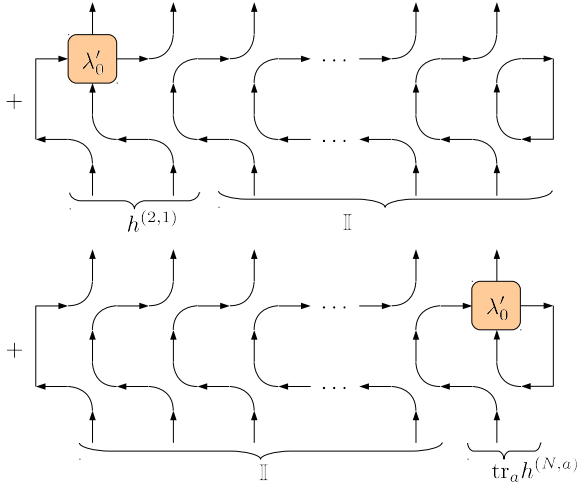}
     \end{center}
    \caption{
	Derivation of the open boundary condition Hamiltonian by
	derivative of the transfer matrix shown in Fig.~\ref{fig:fundamentalrepopenbc2}b
	at the point $\lambda_0$.
	(part~II).
       }
    \label{fig:hamiltonianopenbc2}
\end{figure}

The transfer matrix constructed in this way is indeed related to a local Hamiltonian
with open boundary conditions.
This Hamiltonian is obtained as the derivative of the transfer matrix at the point
$\lambda_0$ at which the $R$-tensor is equal to the identity (see~(\ref{eqn:regularity})).
Explicitly, the obtained Hamiltonian is of the form
\begin{equation} \label{eqn:hamiltonianopenbc}
H \equiv \sum_{i=1}^{N-1} h^{(i,i+1)} + \frac{1}{d} \tr_a h^{(N,a)}
\end{equation}
and related to the transfer matrix via
\begin{displaymath}
H = \frac{1}{2 d} \tau'(\lambda_0).
\end{displaymath}
In~(\ref{eqn:hamiltonianopenbc}), the symbol $a$ refers to an auxiliary system that is traced out.
Using the notation from appendix~\ref{sec:bethe}, this relation is seen as follows:
the derivative $\tau'(\lambda_0)$ disintegrates into a sum of $2 N$ terms, each term
containing one tensor differentiated at~$\lambda_0$ and $2N-1$ tensors evaluated at~$\lambda_0$.
Due to the regularity condition~(\ref{eqn:regularity}) of the $R$-tensor,
the tensors evaluated at~$\lambda_0$ assume the simple form shown in Fig.~\ref{fig:hamiltonian}a.
As can be gathered from Figs.~\ref{fig:hamiltonianopenbc} and~\ref{fig:hamiltonianopenbc2},
each differentiated tensor at site~$i$ corresponds to a two-site term~$h^{(i,i+1)}$ for $i=1,\ldots,N-1$
(with $h$ being defined in~(\ref{eqn:defh})). For $i=N$, two indices of the
tensor are traced out, which leads to the one-site term $\tr_a h^{(N,a)}$.

\subsection{Bethe Ansatz for the XXZ model with open boundary conditions}

For the XXZ model, the virtual dimension~$d$ is equal to~$2$, such that the
Monodromy can be written in the form
\begin{displaymath}
\mathcal{K}^-(\lambda) = 
\left( 
\begin{array}{cc} \mathcal{A}(\lambda) & \mathcal{B}(\lambda)\\ \mathcal{C}(\lambda) & \mathcal{D}(\lambda)
\end{array}
\right)
\end{displaymath}
with
\begin{displaymath}
\begin{array}{ccclccc}
\mathcal{A}(\lambda) & = & \mathcal{K}^-_{0 0}(\lambda), & & \mathcal{C}(\lambda) & = & \mathcal{K}^-_{1 0}(\lambda)\\
\mathcal{B}(\lambda) & = & \mathcal{K}^-_{0 1}(\lambda), & & \mathcal{D}(\lambda) & = & \mathcal{K}^-_{1 1}(\lambda).
\end{array}
\end{displaymath}
The $R$-tensor has the form~(\ref{eqn:Rgl2}) with $b(\lambda)$ and
$c(\lambda)$ being defined by~(\ref{eqn:RXXZb}) and~(\ref{eqn:RXXZc}).
It fulfills the regularity condition~(\ref{eqn:regularity}) at the point $\lambda_0 = 0$,
the unitarity condition~(\ref{eqn:unitarityopenbc}) with $\rho(\lambda)=1$ and
the crossing unitarity condition~(\ref{eqn:crossingunitarityopenbc})
with $c=2 i \eta$ and
$\tilde{\rho}(\lambda) = 1 - \sin(2\eta)^2/\sin(2\eta - i \lambda)^2$.
Using representation~(\ref{eqn:krepopenbc}) for $\mathcal{K}^-(\lambda)$, the $R$-tensor
generates the Hamiltonian
\begin{displaymath}
H = \frac{1}{\sinh(2 i \eta)} (H_{XXZ}^{obc}(\Delta) - \Delta)
\end{displaymath}
with
\begin{displaymath}
H_{XXZ}^{obc}(\Delta) = \sum_{n=1}^{N-1} h_{XXZ}(\Delta).
\end{displaymath}

The precondition for the Bethe Ansatz is that a representation must exist, for which there is a
pseudo-vacuum~$\ket{vac}$ that is an eigenstate of
$\mathcal{A}(\lambda)$ and $\mathcal{D}(\lambda)$ and that is annihiliated by $\mathcal{C}(\lambda)$:
\begin{eqnarray*}
\mathcal{A}(\lambda) \ket{vac} & = & a(\lambda) \ket{vac}\\
\mathcal{D}(\lambda) \ket{vac} & = & d(\lambda) \ket{vac}\\
\mathcal{C}(\lambda) \ket{vac} & = & 0.
\end{eqnarray*}
As argumented before, the operator $\mathcal{C}(\lambda)$ annihilates one down-spin,
whereas $\mathcal{A}(\lambda)$ and $\mathcal{D}(\lambda)$ keep the number of down-spins constant,
such that the state with all spins up is a valid pseudo-vacuum.
The goal is now to diagonalize the transfer matrix
$\tau(\lambda) = \mathcal{A}(\lambda) + \mathcal{D}(\lambda)$.
Since all transfer matrices commute, $[\tau(\lambda),\tau(\mu)]=0$,
all eigenvectors are independent of $\lambda$. The eigenvalue problem reads
\begin{displaymath}
\tau(\lambda) \ket{\Psi} = \tau(\lambda) \ket{\Psi}.
\end{displaymath}
The Bethe Ansatz
\begin{displaymath}
\ket{\Psi(\mu_1,\ldots,\mu_M)} = \mathcal{B}(\mu_1) \cdots \mathcal{B}(\mu_M) \ket{vac}.
\end{displaymath}
fulfills the eigenvalue problem provided that the $\mu_j$'s fulfill the
Bethe equations.
The proof is based upon the algebraic relations between $\mathcal{A}(\lambda)$, $\mathcal{B}(\lambda)$,
$\mathcal{C}(\lambda)$ and $\mathcal{D}(\lambda)$ and is described in detail in~[\onlinecite{sklyanin88}].

Defining the momenta $p_j$ via
\begin{displaymath}
p_j = i \ln \frac{\mu_j}{\mu_j+\eta},
\end{displaymath}
the Bethe equations in their logarithmic form read~\cite{gaudin71,alcaraz87}
\begin{displaymath}
(N+1) p_n = \pi I_n + \Theta(p_n,-p_n) +
\sum_{\begin{smallmatrix} j=1\\ j\neq n \end{smallmatrix}}^M
\frac{\Theta(p_n,-p_j) + \Theta(p_n,p_j)}{2}
\end{displaymath}
with $\Theta(p,q)$ being defined in~(\ref{eqn:deftheta}).
The ground state for $N$ even and $M=N/2$ corresponds to $(I_1,\ldots,I_M) = (1,3,\ldots,N-1)$.
The energy eigenvalue of $H_{XXZ}^{obc}(\Delta)$ for a configuration $(p_1,\ldots,p_M)$ is obtained as
\begin{displaymath}
E_{XXZ}^{obc}(\Delta) = - 2 \sum_{j=1}^M ( \Delta - \cos(p_j) ).
\end{displaymath}


\section{Conclusions}

Summing up, we have sketched the Algebraic Bethe Ansatz using the
pictoresque language of tensor networks.
\emph{In a future paper, the method will be extended to [three-dimensional]
space lattices and its physical implications for cohesion, ferromagnetism
and electrical conductivity will be derived.}~\cite{bethe31}


\acknowledgments{
V.~M.\ and F.~V.\ acknowledge support from the SFB projects
FoQuS and ViCoM, the European projects Quevadis, and the ERC grant
Querg.
V.~K achnowledges support from the NSF grant Grant DMS-0905744.
}


\appendix


\section{Algebraic Derivation of the Bethe Equations} \label{app:derivbethe}

For completeness, we sketch here the derivation of the Bethe Equations
using algebraic relations. We thereby follow Korepin~\cite{korepin93}.

The goal is to find eigenvectors of $t(\lambda) = A(\lambda) + D(\lambda)$
using algebraic relations between $A(\lambda),B(\lambda),C(\lambda)$ and $D(\lambda)$.
The commutation relations that are required are
\begin{eqnarray}
B(\lambda) B(\mu) = & B & ( \mu) B(\lambda) \label{eqn:app:commbb} \\
A(\lambda) B(\mu) = & f & ( \lambda,\mu) B(\mu) A(\lambda) + g(\lambda,\mu) B(\lambda) A(\mu) \label{eqn:app:commab} \\
D(\lambda) B(\mu) = & f & ( \mu,\lambda) B(\mu) D(\lambda) + g(\mu,\lambda) B(\lambda) D(\mu) \label{eqn:app:commad}
\end{eqnarray}
Here, the abbreviations $f(\lambda,\mu) = 1/c(\mu,\lambda)$ and
$g(\lambda,\mu) = - b(\mu,\lambda)/c(\mu,\lambda)$ are used.

The Bethe Ansatz reads
\begin{equation} \label{eqn:app:bethe}
\ket{\Psi(\mu_1,\ldots,\mu_M)} = B(\mu_1) \cdots B(\mu_M) \ket{vac},
\end{equation}
where 
$\ket{vac}$ is a state that is an eigenvector of $A(\lambda)$ and $D(\lambda)$ with
eigenvalues $a(\lambda)$ and $d(\lambda)$, and that is annihilated by $C(\lambda)$.
$A(\lambda)$ applied to $\ket{\Psi(\mu_1,\ldots,\mu_M)}$ using relation~(\ref{eqn:app:commab}) yields
in principle~$2^M$ terms, because each commutation of $A(\lambda)$ with a $B(\mu_k)$ yields $2$ terms
and it takes $M$ commutations to move $A(\lambda)$ from left to right.
However, these two terms are not arbitrary. Both terms only perform exchange operators:
the $f$-term in~(\ref{eqn:app:commab}) exchanges the operators $A$ and $B$, \emph{but not} their arguments;
the $g$-term, on the other hand, exchanges the operators $A$ and $B$ \emph{and} their arguments.
Due to this, after $M$ commutations the following conditions must hold:
\begin{itemize}
\item Every term must contain $M$ $B$'s and one $A$.
\item The $M+1$ coefficients $(\lambda,\mu_1,\ldots,\mu_M)$ are distributed among the $M$ $B$'s and the one $A$.
\end{itemize}
Since all $B$'s commute, there are only~$2$ cases: either $\lambda$ is argument of $A$ -- then the term looks like
\begin{equation} \label{eqn:app:term1}
B(\mu_1) \cdots B(\mu_M) A(\lambda) \ket{vac}.
\end{equation}
Or, $\lambda$ is argument of one of the $B$'s. Then the term is of the form
\begin{equation} \label{eqn:app:term2}
B(\lambda) \prod_{j \neq n} B(\mu_j) A(\mu_n) \ket{vac}
\end{equation}
with $n \in \{1,\ldots,M\}$. Thus, the $2^M$ terms can be collected into $M+1$ linearly independent terms:
\begin{eqnarray*}
A(\lambda) \ket{\Psi(\mu_1,\ldots,\mu_M)} =
 & \Lambda & B(\mu_1) \cdots B(\mu_M) A(\lambda) \ket{vac} \\
+ \sum_{n=1}^M & \Lambda_n & B(\lambda) \prod_{j \neq n} B (\mu_j) A(\mu_n) \ket{vac}
\end{eqnarray*}
What remains to be done is the calculation of the coefficients $\Lambda$ and $\Lambda_n$. 

The expression~(\ref{eqn:app:term1}) is obviously obtained after $M$ commutations 
using the $f$-term in~(\ref{eqn:app:commab}). The $g$-term must not be applied, because
it introduces a $B(\lambda)$. Thus
\begin{displaymath}
\Lambda = \prod_{j=1}^M f(\lambda,\mu_j).
\end{displaymath}
To obtain~(\ref{eqn:app:term2}), it is convenient to rewrite the Bethe Ansatz~(\ref{eqn:app:bethe}) as
\begin{displaymath}
\ket{\Psi(\mu_1,\ldots,\mu_M)} = B(\mu_n) \prod_{j \neq n} B(\mu_j) \ket{vac}.
\end{displaymath}
This is possible for all~$n$, since all $B$'s commute.
Since expression~(\ref{eqn:app:term2}) must not contain $B(\mu_n)$, the first commutation with $A(\lambda)$
must be performed using the $g$-term in~(\ref{eqn:app:commab}).
The expression then reads
\begin{displaymath}
g(\lambda,\mu_n) B(\lambda) A(\mu_n) \prod_{j \neq n} B(\mu_j) \ket{vac}.
\end{displaymath}
All further commutations must use the $f$-term, because another use of the $g$-term would
introduce $B(\mu_n)$ in the expression again.
Thus, the coefficients must be
\begin{displaymath}
\Lambda_n = g(\lambda,\mu_n) \prod_{j \neq n} f(\mu_n,\mu_j)
\end{displaymath}

The application of $D(\lambda)$ to $\ket{\Psi(\mu_1,\ldots,\mu_M)}$ can be
treated in a similar way using relations~(\ref{eqn:app:commbb}) and (\ref{eqn:app:commad}).
Again, the application yields $M+1$ terms
\begin{eqnarray*}
D(\lambda) \ket{\Psi(\mu_1,\ldots,\mu_M)} =
 & \tilde{\Lambda} & B(\mu_1) \cdots B(\mu_M) D(\lambda) \ket{vac} \\
+ \sum_{n=1}^M & \tilde{\Lambda}_n & B(\lambda) \prod_{j \neq n} B (\mu_j) D(\mu_n) \ket{vac}
\end{eqnarray*}
The coefficients are
\begin{displaymath}
\tilde{\Lambda} = \prod_{j=1}^M f(\mu_j,\lambda).
\end{displaymath}
and
\begin{displaymath}
\tilde{\Lambda}_n = g(\mu_n,\lambda) \prod_{j \neq n} f(\mu_j,\mu_n).
\end{displaymath}

Thus, $\ket{\Psi(\mu_1,\ldots,\mu_M)}$ is an eigenvector of $t(\lambda) = A(\lambda) + D(\lambda)$
if
\begin{displaymath}
a(\mu_n) \Lambda_n + d(\mu_n) \tilde{\Lambda}_n = 0
\end{displaymath}
for $n=1,\ldots,M$. These relations are the Bethe Ansatz equations, which can be
written in the form
\begin{displaymath}
\frac{d(\mu_n)}{a(\mu_n)} =
\prod_{\begin{smallmatrix} j=1\\ j\neq n \end{smallmatrix}}^M
\frac{c(\mu_n,\mu_j)}{c(\mu_j,\mu_n)}
\end{displaymath}
under the assumption that $g(\lambda,\mu)$ is an odd function in the sense that $g(\lambda,\mu)=-g(\mu,\lambda)$
(as it is the case for the Heisenberg model and the XXZ model).

The eigenvalue $\tau(\lambda)$ is obtained as
\begin{displaymath}
\tau(\lambda) = a(\lambda) \Lambda + d(\lambda) \tilde{\Lambda},
\end{displaymath}
which can be expressed as
\begin{displaymath}
\tau(\lambda) =
a(\lambda) \prod_{j=1}^M \frac{1}{c(\mu_j,\lambda)} +
d(\lambda) \prod_{j=1}^M \frac{1}{c(\lambda,\mu_j)}.
\end{displaymath}


%


\end{document}